\documentclass[11pt,a4paper]{article}
\usepackage{jheppub_r}
\usepackage{enumerate}
\usepackage{afterpage}

\newcommand{\nc}{\newcommand}

\newcommand{\rdec}{r_\mathrm{dec}}
\newcommand{\ba}{\begin{eqnarray}}
\newcommand{\ea}{\end{eqnarray}}
\newcommand{\Nrel}{N_\mathrm{rel}}
\newcommand{\Ndec}{N_\mathrm{dec}}

\newcommand{\hf}{\frac{1}{2}}
\newcommand{\bea}[1]{\begin{eqnarray} \mbox{$\label{#1}$}}
\newcommand{\eea}{\end{eqnarray}}
\newcommand{\be}[1]{\begin{equation} \mbox{$\label{#1}$}}
\newcommand{\ee}{\vspace{0.1cm}\end{equation}}
\newcommand{\eq}[1]{\mbox{(\ref{#1})}}
\newcommand{\fig}[1]{\mbox{Fig.\ (\ref{#1})}}
\newcommand{\sect}[1]{\mbox{section\ \ref{#1}}}
\newcommand{\nl}{\nonumber \\}
\def\GeV{{\rm \ GeV}}
\def\MeV{{\rm \ MeV}}
\nc{\fb}[2]{\left(\frac{#1}{#2}\right)}
\nc{\sqb}[2]{\sqrt{\frac{#1}{#2}}}
\nc{\fnl}{f_{_{NL}}}

\nc{\Np}{N_\mathrm{pre}}
\nc{\zobs}{\zeta_\mathrm{obs}}
\nc{\meff}{m_{eff}}

\nc{\mpl}{M_{_P}}

\nc{\lambdasigma}{\lambda_\sigma}

\nc{\obs}{{\em observable}}
\nc{\negl}{{\em negligible}}
\nc{\excl}{{\em excluded}}
\nc{\Npre}{N_{pre}}

\title{Quantifying the `naturalness' of the curvaton model}
\author[a,c]{Rose N. Lerner}
\author[c]{and Scott Melville}
\affiliation[a]{Deutsches Elektronen-Synchrotron DESY, 22607 Hamburg, Germany}
\affiliation[c]{Helsinki Institute of Physics, P.O. Box 64, FI-00014, Helsinki, Finland.}

\emailAdd{rose.lerner@desy.de, scott.melville@queens.ox.ac.uk}

\preprint{DESY-14-015}

\abstract{We investigate the probability of obtaining an observable curvature perturbation, using as an example the minimal curvaton-higgs (MCH) model. We determine ``probably observable'' and ``probably excluded'' regions of parameter space assuming generic initial conditions and applying a stochastic approach for the curvaton's evolution during inflation. Inflation is assumed to last longer than the $N_{obs} \simeq 55$ observable $e$-folds, and the {\em total} number of $e$-folds of inflation determines the particular ranges of parameters that are probable. For the MCH model, these ``probably observable'' regions always lie within the range $8\times 10^4 \GeV \leq m_\sigma \leq 2\times 10^7\GeV$, where $m_\sigma$ is the curvaton mass, and the Hubble scale at horizon exit is chosen as $H_* = 10^{10}\GeV$. Because the ``probably observable'' region depends on the total duration of inflation, information on parameters in the Lagrangian from particle physics and from precision CMB observations can therefore provide information about the total duration of inflation, not just the last $N_{obs}$ $e$-folds. This method could also be applied to any model that contains additional scalar fields to determine the probability that these scalar fields contribute to the curvature perturbation.}

\begin{document}
\maketitle

\section{Introduction}
\label{sec:intro}
In order to generate sufficient inflation, and in order to obtain the observed amplitude of scalar perturbations, an inflationary theory must give certain initial conditions to the fields involved. In this paper we focus on this problem as applied only to spectator fields, and study the initial field value of the curvaton field.

The curvaton scenario is an alternative mechanism to generate the primordial curvature perturbation $\zeta$ \cite{curvorig}. It is interesting both because it can allow many inflaton models which are successful for generating inflation but cannot produce sufficient perturbations, and because curvaton models can produce features such as non-Gaussianity and isocurvature, allowing constraints from data to be placed. In addition, the coupling of the curvaton to other fields is less tightly constrained than for the inflaton, allowing for stronger couplings to the standard model.

Although the curvaton model has been well-studied, it is often suggested that the scenario is fine-tuned or unnatural. The parameters of a typical curvaton model are: the Hubble parameter $H_*$, which gives the energy scale of inflation; the effective curvaton decay width $\Gamma$; the parameters of the potential such as the mass $m_\sigma$ and quartic self-coupling $\lambda_\sigma$; and the value of the homogeneous curvaton field during inflation $\sigma_*$, determined as observable scales exit the horizon approximately $N_{obs}\simeq55$ $e$-folds before the end of inflation. Generally, by tuning either $\Gamma$ or $\sigma_*$, the correct amplitude of curvature perturbation can be obtained, $\zeta =4.7\times 10^{-5}$ \cite{Planck:zeta}. Some attempts have been made to determine $\Gamma$ in terms of the particle physics couplings of the model. A specific example is the non-perturbative decay of the curvaton into standard model Higgs bosons \cite{Enqvist:2012tc,Enqvist:2013qba,Enqvist:2013gwf,Mukaida:2014yia}; the perturbative case is discussed in \cite{Enqvist:2011jf}.

In this paper, we investigate the probability of obtaining the measured curvature perturbation in the minimal curvaton-higgs (MCH) model. In this model, the curvaton is a real singlet scalar, coupled to the standard model higgs with the Lagrangian in the unitary gauge given by
\be{Lagrangian}
{\cal L} = {\cal L}_{SM} + \hf \partial_\mu \sigma \partial^\mu \sigma +  \frac{1}{2}m_\sigma^2 \sigma^2 + \frac{\lambdasigma}{4!} \sigma^4 + \frac{1}{4} g^2\sigma^2h^2,
\ee
where ${\cal L}_{SM}$ is the standard model Lagrangian including the higgs and $h$ is the physical higgs boson. Because the higgs has a large field value during inflation, $h \gg 246\GeV$,  the higgs potential is well approximated by
\be{higgs_potential}
V(h) = \frac{\lambda(\mu)}{4} h^4,
\ee
where $\lambda(\mu)$ is the higgs self coupling at the relevant energy scale. We now assume that the curvaton's effective quartic coupling is negligible, i.e.\ $\lambdasigma = 0$.  In general, effective quartic terms are generated by loop corrections\footnote{See \cite{Enqvist:2011jf} for a discussion of this in the curvaton scenario.}, however, setting $\lambdasigma = 0$ enables us to calculate the stochastic behaviour (mostly) analytically.

We assume that inflation lasts for a total duration of $N = N_{pre} + N_{obs}$ $e$-folds. We then use a stochastic approach \cite{Starobinsky86} to quantify the probability of obtaining particular values of $\sigma_*$, calculated $N_{obs}$ $e$-folds before inflation ends. Because the observables $\zeta$ (amplitude of the curvature perturbation) and $\fnl$ (non-Gaussianity) depend on $\sigma_*$, we can then translate the probability distribution of $\sigma_*$ into a probability for obtaining any particular range of these observables. This particular approach was first suggested in \cite{Enqvist:2012xn}, although applying the equilibrium limit of the stochastic dynamics to the curvaton model was already considered in \cite{Lyth:FP}. The influence of superhorizon perturbations on observables has also been discussed, e.g.\ in \cite{superhor}.

We find that for most values of $N_{pre}$ there are wide ranges of $g$ and $m_\sigma$ that give a high ($>10\%$) probability of producing a curvature perturbation that is \obs,\ i.e.\ neither too large to be excluded, nor too small to be negligible. This is significant because it enables us to relate the parameters of the model to knowledge about the duration of inflation, which could be found from a fundamental theory. Conversely, information from particle physics and CMB observations could give us information about the duration of inflation before the last $N_{obs}$ $e$-folds, via their predictions for curvaton model parameters.

This methodology could also be used to quantify, under various assumptions, whether additional light scalars in any particular model would be likely to give a non-negligible contribution to predictions.

The paper is organised as follows. In \sect{sec:dyn} we introduce the relevant stochastic evolution equation, discuss the decay mechanisms for the curvaton field, and calculate the contribution of the higgs field during inflation to the curvaton's effective mass. In \sect{sec:prob} we calculate the probability that the model gives $\zeta$ and $\fnl$ that are either \obs,\ \negl\ or \excl\ by current observations. Then in \sect{sec:results} we present our results first for the case $N_{pre} \to \infty$ and secondly for the case of finite $N_{pre}$. Finally in \sect{sec:conc} we conclude.

\section{Dynamics of the spectator field}
\label{sec:dyn}

In this section we first introduce and solve the Fokker-Planck equation describing the stochastic dynamics of the curvaton field, then we review the relevant decay mechanisms in the MCH scenario, and finally we address the effect of the higgs on the curvaton during inflation.

\subsection{Stochastic dynamics}

During inflation, the curvaton is a light spectator field with an effective mass $\meff < H_*$. It typically experiences both quantum fluctuations and a classical slow roll towards its minimum. The long wavelength behaviour of spectator fields in a de Sitter background can be described using a stochastic approach \cite{Starobinsky86,Starobinsky} (see also e.g.~\cite{oldstochastic} for some early references on the subject, and \cite{Enqvist:2012xn} for recent numerical work).

To calculate the stochastic behaviour of the spectator $\sigma$, we assume that the universe is de Sitter, that the Hubble rate remains constant throughout inflation, and that the curvaton does not dominate the energy density during inflation. The mechanism of inflation is unimportant provided that the Hubble parameter can be approximated as constant during inflation, $H(t)\simeq H_*$. We also assume the absence of couplings between the curvaton and the inflaton, which could affect the dynamics during inflation\footnote{In general, the inflaton requires small coupling in order for its potential to remain sufficiently flat.}.

The stochastic behaviour is described by a Fokker-Planck equation, which is obtained by integrating out the short wavelength modes --- see e.g.\ \cite{Starobinsky} for details. Throughout this paper, we use the number of $e$-folds $N$ as a time scale. $P(\sigma,N)$ represents the probability that at a particular time $N$, the patch that becomes our Universe has a homogeneous curvaton value given by $\sigma$. The resulting Fokker-Planck equation for $P(\sigma,N)$ is \cite{Starobinsky}
\bea{FPeq}
\frac{d P(\sigma,N)}{dN} = \frac{1}{3H_*^2}V''(\sigma)P(\sigma,N)+\frac{1}{3H_*^2}V'(\sigma)P'(\sigma,N)+\frac{H_*^2}{8\pi^2}P''(\sigma,N),
\eea
where the prime is a derivative with respect to $\sigma$. Given an initial probability distribution $P(\sigma,0)$, solving (\ref{FPeq}) yields the distribution for all $N$. Formal solutions to \eq{FPeq} as well as a closed analytical form in the limit $N\to \infty$ are known \cite{Starobinsky}. 

\begin{figure}
\begin{center}
\includegraphics[width=0.7\textwidth, angle=0]{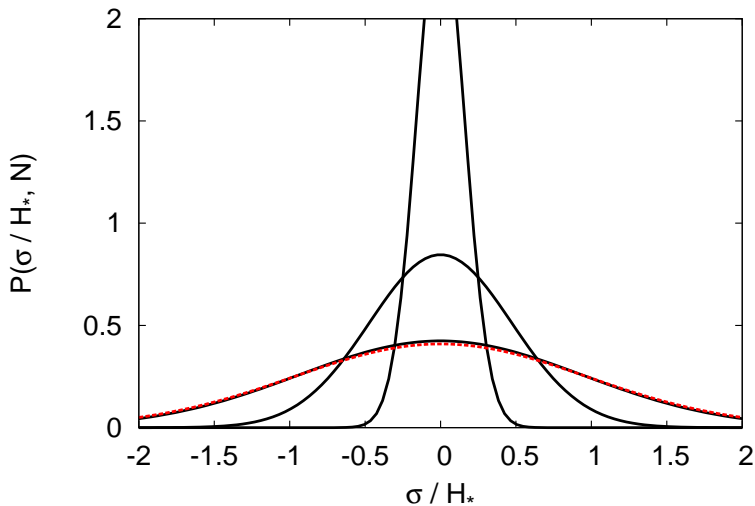} 
\caption{\label{fig:probsigma}
Evolution of $P(\sigma_*,N)$ for a large effective mass, $m_{eff} = 0.2 H_*$, and $\sigma_0 = w_0= 0 $. From narrow to wide: $N_{pre}=1,10,10^2$ (solid), $\Npre = \infty$ (red dashed). For this large $m_{eff}$, the equilibrium limit is already reached at $N_{pre} \simeq {\cal O}(N_{rel}) = 40$.
}
\end{center}
\end{figure}

For finite $N$, \eq{FPeq} should be solved numerically, except in the case of a purely quadratic potential, which we assume in this paper. In that case, the analytical expressions are  \cite{Enqvist:2012xn} 
\be{q1}
P(\sigma,N)=\frac{1}{\sqrt{2\pi w^2(N)}}\exp{\left(-\frac{(\sigma-\sigma_c(N))^2}{2H_*^2 w^2(N)}\right)},
\ee
with
\bea{q2}
\sigma_c(N)&=&\sigma_0\exp{\left(-\frac{N}{\Nrel}\right)},\\
\label{q2other}
w^2(N)&=&\frac{\Ndec}{4\pi^2}-\left(\frac{\Ndec}{4\pi^2}-w^2_0\right)\exp{\left(-\frac{N}{\Ndec}\right)},
\eea
where $\sigma_0$ is the initial central value of the distribution and $w_0$ the initial width. The relaxation and decoherence time-scales are \cite{Enqvist:2012xn} 
\be{rel}
N_{\rm rel}=\frac{3H_*^2}{\meff^2},\qquad N_{\rm dec}=\frac{3H_*^2}{2\meff^2},
\ee
where $\meff$ is the effective mass of the curvaton during inflation. The normalisation of \eq{q1} ensures that $\int_{-\infty}^\infty P(\sigma/H_*,N) d(\sigma/H_*) = 1$.
\fig{fig:probsigma} illustrates \eq{q1} for fixed $m_{eff}$ and various $N$. The distribution relaxes towards equilibrium, in this case achieved after $N_{pre}\simeq \mathcal{O}(\Nrel)$ $e$-folds. As can be seen from \eq{rel}, smaller masses would reach equilibrium much slower.

\subsection{Influence of the higgs on the curvaton during inflation}
\label{karipaper}
Both the curvaton and the higgs are light fields during inflation. The large curvaton-higgs coupling means that the curvaton will gain an effective mass from the higgs. This depends on the higgs field value during inflation. In principle, the higgs field value also has a distribution given by the solution of a Fokker-Planck equation similar to \eq{FPeq}, where the higgs-curvaton coupling would couple the two Fokker-Planck equations. However, for the purposes of this paper, we assume that the higgs field reaches equilibrium quickly during inflation, and we approximate the higgs field value by the root-mean-squared value of the equilibirum distribution\footnote{Regardless of the initial higgs field value, the higgs becomes light after about 30 $e$-folds of inflation \cite{Enqvist:2013kaa}. After that it evolves relatively fast, reaching its equilibrium distribution in $\Delta N \simeq 85$ $e$-folds for $H_* = 10^{10}\GeV$ \cite{Enqvist:2013kaa}.}. This is a reasonable simplification.

The root-mean-squared value of the higgs field $h_*$ depends on the scale of inflation\footnote{
We should also require the higgs to be stable ($\lambda(\mu) > 0$) up to the inflationary scale. This puts bounds on $H_*$ and $h_*$ (see \cite{Enqvist:2013kaa}).
}, 
$H_*$ \cite{Enqvist:2013kaa}. 
During inflation, the higgs bare mass is negligible and the quartic term dominates the potential. The root-mean-squared higgs field value is given by  \cite{Enqvist:2013kaa}
\be{higgsrms} 
h_* \simeq 0.36 \lambda_*^{-1/4} H_*,
\ee
where $\lambda_*$ is the higgs self-coupling at the inflationary scale. The two-loop running of the higgs coupling in the standard model is known, and gives $\lambda_* = 0.0005$ at the scale $H_* = 10^{10}\GeV$ for best-fit standard model parameters \cite{Degrassi}.

The effective mass for the curvaton during inflation is therefore
\be{meff}
m_{eff}^2 = m_\sigma^2 + g^2 h_*^2.
\ee
For small $g$ or large $m_\sigma$, the dominant contribution to the effective mass during inflation is due to the bare mass, i.e.\  $ g h_* \ll m_\sigma.$ For large $g$ or small $m_\sigma$, the dominant contribution to the effective mass during inflation is due to interaction with the higgs, i.e.\ $ g h_* \gg m_\sigma.$

\subsection{Decay of the curvaton field after inflation}

The decay\footnote{By ``decay'' we mean the process by which the homogeneous curvaton energy density that remains after inflation is dissipated into relativistic quanta.} of the homogeneous curvaton field is an essential process in the curvaton scenario. It is at the point of decay that the inflationary perturbations imprinted in the curvaton field become adiabatic perturbations in all the components of the Universe. For a successful model, this decay must occur sufficiently late such that the relative fraction of curvaton has had time to grow and thus that the curvature perturbation will be large enough.

There are three mechanisms that take energy out of the curvaton field. The first is perturbative interactions with the thermal background. It is known that for larger values of $g\gtrsim (m_\sigma/\mpl)^{1/4}$, these will cause a very fast decay of the homogeneous curvaton, and thus would rule out the model \cite{D'Onofrio:2012qy,Enqvist:2012tc,Mukaida:2014yia}. However, the full calculation of these processes is complicated and involved \cite{BasteroGil,Mukaida,Drewes:2013iaa,Mukaida:2014yia}. Because our aim is to illustrate a method of thinking about curvaton initial conditions, we do not calculate and apply this limit here.

The second mechanism is a non-perturbative decay through broad or narrow resonance. This has been studied in the MCH model \cite{Enqvist:2012tc,Enqvist:2013qba}. However, the details are not crucial for this work because it is known that resonance shuts off before all the energy is transferred out of the oscillating field{\footnote{
The fraction remaining is estimated at 5\% in the curvaton case \cite{Chambers:2009ki}, although a robust calculation would require lattice simulations.} \cite{notefficient}. Provided that some fraction of the curvaton remains after this decay, the exact fraction does not qualitatively affect our conclusions and again we ignore this type of decay.

The third mechanism is decay via dimension-5 non-renormalisable operators. Assuming these are suppressed by the Planck scale $\mpl$, the effective decay width has the form
\be{dim5}
 \Gamma_{dim-5} = {\cal O}(1) \frac{m_\sigma^3}{\mpl^2}.
\ee
Thus, provided that the interactions with the thermal bath are negligible, the effective decay width is given by \eq{dim5} and only depends on the curvaton mass\footnote{Note that the homogeneous higgs field decays very soon after inflation ends and thus makes a negligible contribution to the curvaton's effective mass after inflation.}
 $m_\sigma$. An absolute lower bound on the curvaton mass is given by
$m_\sigma \geq 8\times 10^{4}\GeV$, because the curvaton must decay before a temperature of $4\MeV$ to avoid spoiling the predictions of big bang nucleosynthesis (BBN). This requires $\Gamma_{dim-5}\geq H(T=4\MeV)$, giving the bound above.

\section{Defining ``probable'' Universes}
\label{sec:prob}

Different patches of the universe correspond to different realisations of the probability distribution function $P(\sigma,N)$. Thus, our observable Universe has one particular value of $\sigma_*$, drawn from the distribution of $\sigma_*$ that we calculate using the stochastic approach.

\subsection{The curvature perturbation}

In the curvaton model, the homogeneous curvaton begins to oscillate some time after inflation has ended. The background energy density falls $\propto a^{-4}$, whereas the curvaton energy density only falls $\propto a^{-3}$, because it has a quadratic potential. Thus, the relative fraction of curvaton grows. The curvaton decays when $H(t) \simeq \Gamma_{dim-5}$.

The curvature perturbation $\zeta$ is given by a simple expression in the case of a pure quadratic potential \cite{curvorig},
\be{zeta}
\zeta= r_{dec}  \frac{H_*}{3\pi\sigma_*},
\ee
where the relative fraction of curvaton at decay is
\bea{rdec}
r_{dec} &\equiv& \left.\frac{3\rho_\sigma}{3\rho_\sigma+4\rho_{r}}\right|_{decay}\nl
& = & \left(1+8 \frac{m_\sigma \mpl}{\sigma_*^2}\right)^{-1}.
\eea
Each value of $\zeta$ can be obtained with two distinct values of $\sigma_*$ (see \fig{fig:twosigma}). The smaller $\sigma_*$ corresponds to $\rdec < 1$, i.e.\ the curvaton does not entirely dominate the Universe before it decays, whereas the larger $\sigma_*$ corresponds to $\rdec\simeq 1$ i.e.\ a Universe dominated by the curvaton before it decays. For fixed $m_\sigma$ and $H_*$ there is a maximum $\zeta$ that can be obtained, given by
\be{eq:zmax}
\zeta_{max} 
= \sqrt{\frac{H_*^{2}}{288\pi^{2}\mpl m_\sigma}}
\ee
(see also \fig{fig:twosigma}).

\begin{figure}
\begin{center}
\includegraphics[angle=0,width=0.7\textwidth, angle=0]{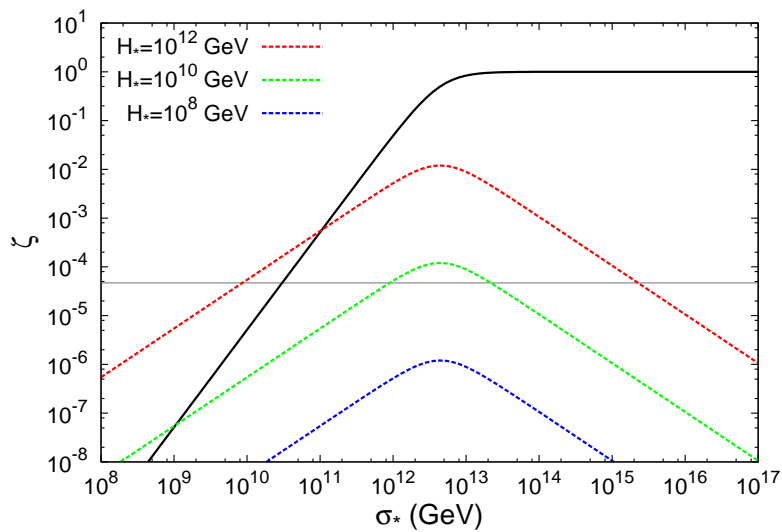} 
\caption{\label{fig:twosigma}
Showing $\zeta(\sigma_*)$ for (dashed, top to bottom) $H_* = 10^{12}\GeV,~10^{10}\GeV$ and $10^8\GeV$, assuming $m_\sigma = 10^{6}\GeV \gg gh_*$. The horizontal line marks the observed  $\zeta_{obs} = 4.7 \times  10^{-5}$. Also shown is the dependence of  $\rdec$ on $\sigma_*$ (solid).
}
\end{center}
\end{figure}

\begin{figure}
\begin{center}
\includegraphics[width=0.5\textwidth, angle=270]{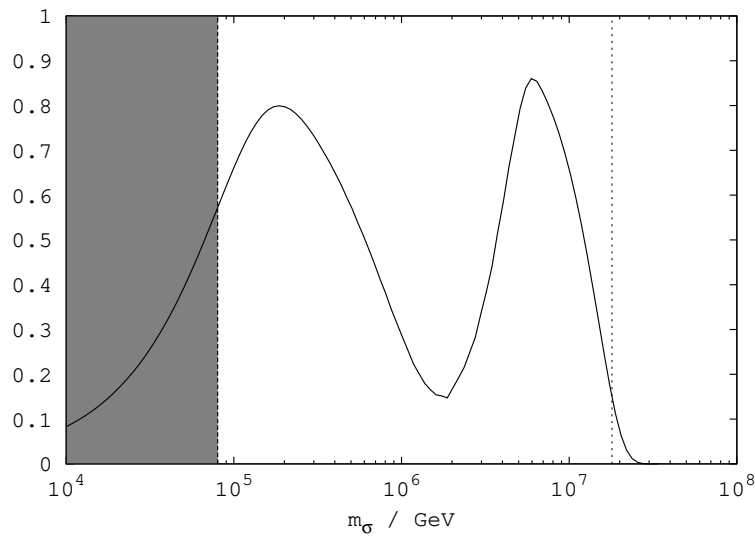} 
\caption{\label{fig:baremass}
Probability of obtaining observable $0.1 \leq \zeta/\zobs \leq 1$ with $\fnl < 14.3$ in the equilibrium limit ($N\to \infty$) for $g h_* \ll m_\sigma$. The shaded region is excluded by the BBN constraint and the dotted line shows how this excluded region would be extended for WIMP dark matter that decouples at $T = 10\GeV$. The peak at low $m_\sigma$ is due to a curvaton with $\rdec \simeq 1$; the high $m_\sigma$ peak is due to a curvaton with $\rdec \lesssim 1$. Plotted for $H_* = 10^{10}\GeV$.
}
\end{center}
\end{figure}

The probability distribution of $\zeta$ is given by
\be{g1}
P(\zeta,N) = P(\sigma_*^{-},N)\left|\frac{d\sigma_*}{d\zeta}\right|_{\sigma_*^{-}} + P(\sigma_*^{+},N)\left|\frac{d\sigma_*}{d\zeta}\right|_{\sigma_*^{+}},
\ee
where $\sigma_*^{-}$ is the smaller value of $\sigma_*$ that gives $\zeta$, and $\sigma_*^{+}$ is the larger. These are given by
\be{sigpm}
\sigma_*^\pm = \frac{H_*}{6\pi\zeta} \left[1 \pm \sqrt{1 - \frac{\zeta^2}{\zeta_{max}^2}}\right] = \frac{H_*}{6\pi\zeta} \left[1 \pm Y(\zeta) \right]\;\;\;\; \mbox{for}\;\;\;\; \zeta \leq \zeta_{max},
\ee
where
\be{Ydef}
Y(\zeta) \equiv \sqrt{1- \frac{\zeta^2}{\zeta_{max}^2}}.
\ee
Note carefully that $m_\sigma$ appears in \eq{g1} via $\zeta_{max}$, and that $\meff$ appears inside the contributions $P(\sigma_*^\pm)$. We thus obtain
\bea{eq:P(zeta)NonEquilibrium}
P(\zeta, N)=&\frac{1}{\sqrt{2\pi w^{2}(N)}}\exp\left(-\frac{\left(\left[ \frac{H_*}{6\pi\zeta}\left(1 - Y(\zeta) \right) \right]-\sigma_c(N) \right)^{2}}{2H_*^{2}w^{2}(N)}\right)\frac{H_*(1-Y(\zeta))}{6\pi\zeta^{2}Y(\zeta)}\nl
&{} + \frac{1}{\sqrt{2\pi w^{2}(N)}}\exp\left(-\frac{\left(\left[ \frac{H_*}{6\pi\zeta}\left(1 + Y(\zeta)\right) \right]-\sigma_c(N)\right)^{2}}{2H_*^{2}w^{2}(N)}\right)\frac{H_*(1+Y(\zeta))}{6\pi\zeta^{2}Y(\zeta)},
\eea
where $w(N)$ and $\sigma_{c}(N)$ are defined in \eq{q2} and \eq{q2other}. The initial conditions $\sigma_0$ and $\omega_0$ are irrelevant for the $\Np \to \infty$ case.
Note that \eq{eq:P(zeta)NonEquilibrium} is only valid for $\zeta \leq \zeta_{max}$, where $\zeta_{max}$ is given by \eq{eq:zmax}. Correct normalisation implies that $\int_0^{\zeta_{max}} P(\zeta,N) d\zeta = 1$. 

We are interested in all cases where the curvaton makes an observable contribution to the curvature perturbation: we define this range to be $0.1 \leq \zeta/\zobs \leq 1$. This allows for a mixed inflaton-curvaton scenario where part of the perturbation comes from the inflaton.

\fig{fig:baremass} illustrates the probability of obtaining $0.1 \leq \zeta/\zobs \leq 1$ and $\fnl<14.3$ in the equilibrium limit ($N\to \infty$). Two peaks are visible. These are due to the curvaton with $\rdec \simeq 1$ (low $m_\sigma$) and the curvaton with $\rdec < 1$ (higher $m_\sigma$). The figure should be interpreted as follows. For each independent $m_\sigma$, the probability of obtaining an ``observable'' $\zeta$ is shown, obtained by integrating \eq{eq:P(zeta)NonEquilibrium} with respect to $\zeta$.

\subsection{Non-Gaussianity $\fnl$}
The non-gaussianity of the primordial perturbation is well constrained by Planck in the case of local non-Gaussianity to be $\fnl \leq 14.3$ at 2-sigma \cite{Planck:NG}. In the case of $\rdec \simeq 1$, the non-Gaussianity is $\fnl = {\cal O}(1)$ and is thus not constrained by observations. However, in the limit $\rdec \ll 1$, $\fnl$ can be large.

The inflaton is assumed to contribute negligible non-Gaussianity. Also assuming uncorrelated inflaton and curvaton perturbations, in the $\delta N$ formalism, $\zeta$ and the total amount of non-Gaussianity $\fnl^{^{total}}$ are given by \cite{NG}
\be{deltaN_zeta}
\zeta = N_\sigma \delta \sigma + N_\phi \delta \phi + \cdots,
\ee
and
\be{deltaN_fnl}
\frac{6}{5}\fnl^{^{total}} = \frac{N_\sigma^2 N_{\sigma \sigma}}{(N_\sigma^2 + N_\phi^2)^2},
\ee
where subscripts denote derivatives.
The non-Gaussianity from the curvaton perturbation is
\be{deltaN_curv}
\frac{6}{5}\fnl^\sigma = \frac{N_{\sigma\sigma}}{N_\sigma^2} \simeq \frac{3}{2\rdec},
\ee
where the approximation is good for a quadratic potential. Allowing for the case where the curvaton only contributes some fraction $\zeta^\sigma/\zobs$ of the observed $\zeta$, we find  
\bea{fnl_1}
\fnl^{^{total}} &=& \fb{\zeta^\sigma}{\zobs}^4 \fnl^\sigma \nl
& \simeq & \frac{5}{4} \fb{1}{3\pi \zeta_{obs}}^4 \, \fb{H_*}{\sigma_*}^4\,\, \rdec^3(\sigma_*).
\eea
The probability of obtaining a particular $\fnl^{^{total}}$ is then given either by
\bea{fnl_3}
P(f_A < \fnl^{^{total}} < f_B) &= & \int^{f_B}_{f_A} d\fnl \,\, P[\sigma_*(\fnl)]\left| \frac{d\sigma_*}{d\fnl} \right|_{\sigma_*(\fnl)} 
\eea
where $\sigma_*(\fnl)$ is the value of $\sigma_*$ corresponding to $\fnl^{^{total}}$, obtained by solving \eq{fnl_1}. Equivalently the probability is given by
\bea{late}
P(f_A < \fnl < f_B) &= &
\int_{\zeta(f_A)}^{\zeta(f_B)} d\zeta \,\, P[\sigma_*^{-}]\left|\frac{d\sigma_*}{d\zeta}\right|_{\sigma_*^{-}} + \int_{\zeta(f_A)}^{\zeta(f_B)} d \zeta\,\, P[\sigma_*^{+}]\left|\frac{d\sigma_*}{d\zeta}\right|_{\sigma_*^{+}},
\eea
noting that the second integral only picks out the value $\fnl^{\sigma} = 5/4$ in our approximations.

Because we want to combine limits on $\zeta$ and $\fnl^{^{total}}$, we will use the second method, \eq{late}. In practice our integration limits will be max($\zeta(f_A), 0.1\zeta_{obs}$) and min($\zeta(f_B),\zeta_{obs}$). For completeness, we give $\zeta(f_A)$, which is
\be{zeta_\fnl}
\zeta(\fnl^{^{total}}) = 12\pi  \zeta_{obs}^2 \sqrt{2/5} \sqrt{m_\sigma \mpl/H_*^2} \sqrt{\fnl^{^{total}}},
\ee
where we have used $r_{dec} \ll 1$ in the derivation, an approximation that can add 10\% error.

\section{Results}
\label{sec:results}

This section displays plots for various values of $\Npre$, scanning over $g$ and $m_\sigma$. In each case, we divide the parameter space into \obs,\ \negl,\ and \excl\ regions as follows:
\begin{enumerate}
 \item {\bf Observable:} {\bf either (a)} the curvaton contributes at least 10\% and not more than 100\% of $\zeta_{obs}$, and $\fnl$ is below observational bounds, $4.7\times 10^{-6} \leq \zeta \leq 4.7\times 10^{-5}$ and $\fnl < 14.3$. {\bf or (b)} the curvaton's contribution to $\zeta$ is negligible but it still gives large non-Gaussianity, $\zeta < 4.7\times 10^{-6}$ and $5 \leq \fnl \leq 14.3$.
 \item {\bf Negligible:} The curvaton contributes negligibly to $\zeta$ and the non-Gaussianity is below the expected Planck sensitivity, $\zeta < 0.1\zeta_{obs}$ and $\fnl < 5$.
 \item {\bf Excluded:} {\bf either (a)} The curvaton's contribution to $\zeta$ is too high, $\zeta \geq \zeta_{obs}$. {\bf or (b)} the curvaton gives too-high non-Gaussianity, $\zeta \leq 4.7\times 10^{-5}$ and $\fnl > 14.3$.
\end{enumerate}
We do not and cannot define a strict boundary between ``probable'' and ``improbable''; we do however plot the probability of the model lying in each of the three regions, marking in particular where the probability is greater than 1\% and greater than 10\%. There is therefore some overlap of regions.

It is important and interesting to consider all three regions if we wish to investigate the impact of adding new scalars to the model and not just whether a particular scalar lies in the \obs\ range. In the \negl\ region, a scalar with generic initial conditions could exist without having any impact on CMB observables. However, in the \excl\ region, a scalar would need fine-tuned initial conditions in order to be allowed.

The too-large $\fnl$ case ((b) of \excl) is a small part of the \excl\ region, but is nevertheless important to include. In contrast, we find analytically and numerically that the case with small $\zeta$ but large $\fnl$ ((b) of \obs) can never occur. It would require $\zeta_{max} > \zeta(\fnl = 5)<0.1\zobs$. The second of these conditions is only satisfied for $m_\sigma$ below the BBN lower limit (for $H_* = 10^{10}\GeV$).

We have checked that the total probability of the three regions adds up to 1. We have also made rough analytical checks by computing a typical value of $\sigma_*$ for various values of $m_\sigma \gg gh_*$ from the distribution\footnote{We do this by setting the argument of the exponential in $P(\sigma,N=\infty)$ equal to 1.}. Analytically computing $\zeta$ using \eq{zeta} then gives values of $\zeta$ that lie within \obs,\ \negl\ and \excl\ as appropriate.

\subsection{Equilibrium: $N \to \infty$}
\afterpage{\clearpage}
\begin{figure}
\begin{center}
\includegraphics[width=0.45\textwidth, angle=270]{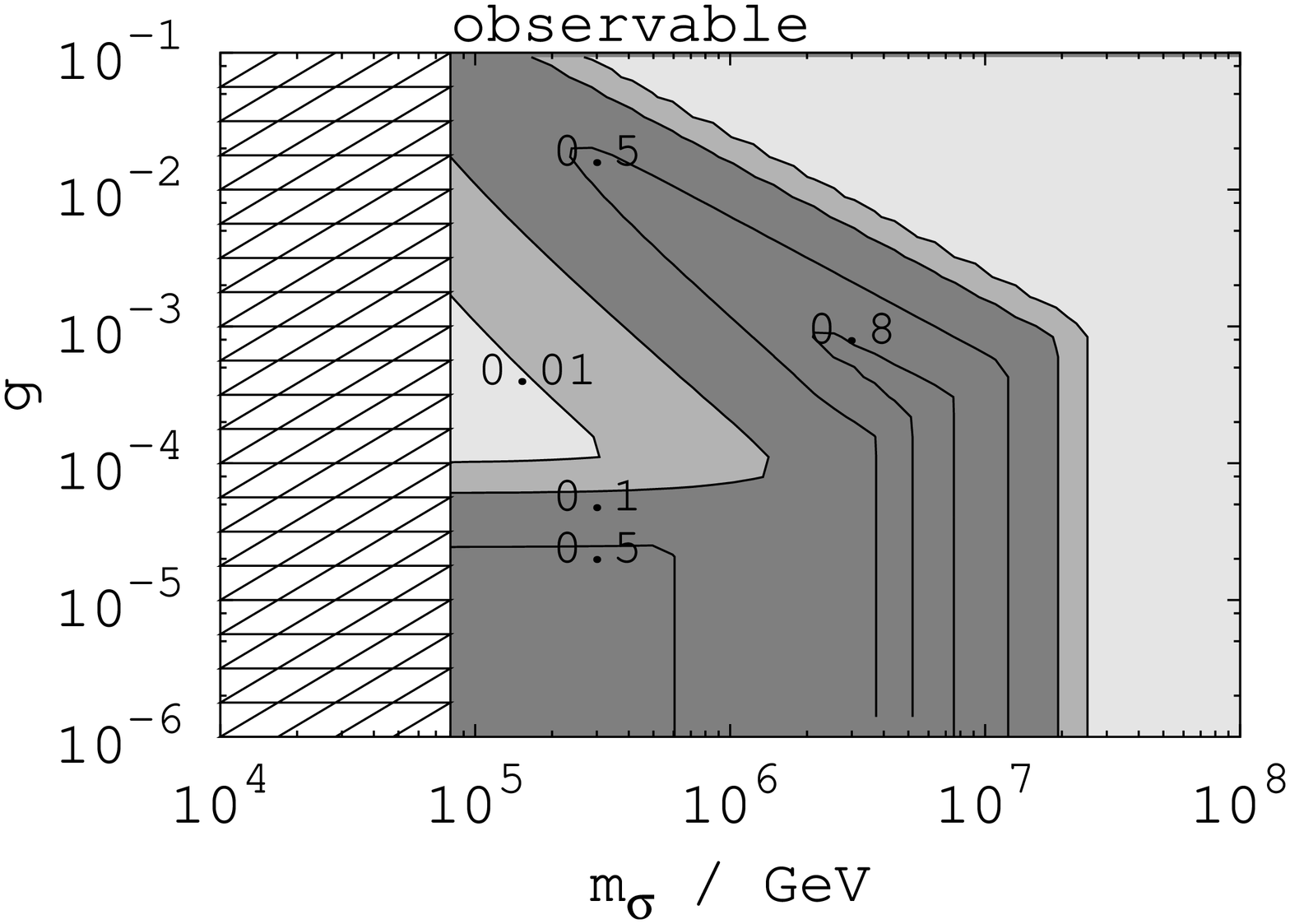}\\
\includegraphics[width=0.45\textwidth, angle=270]{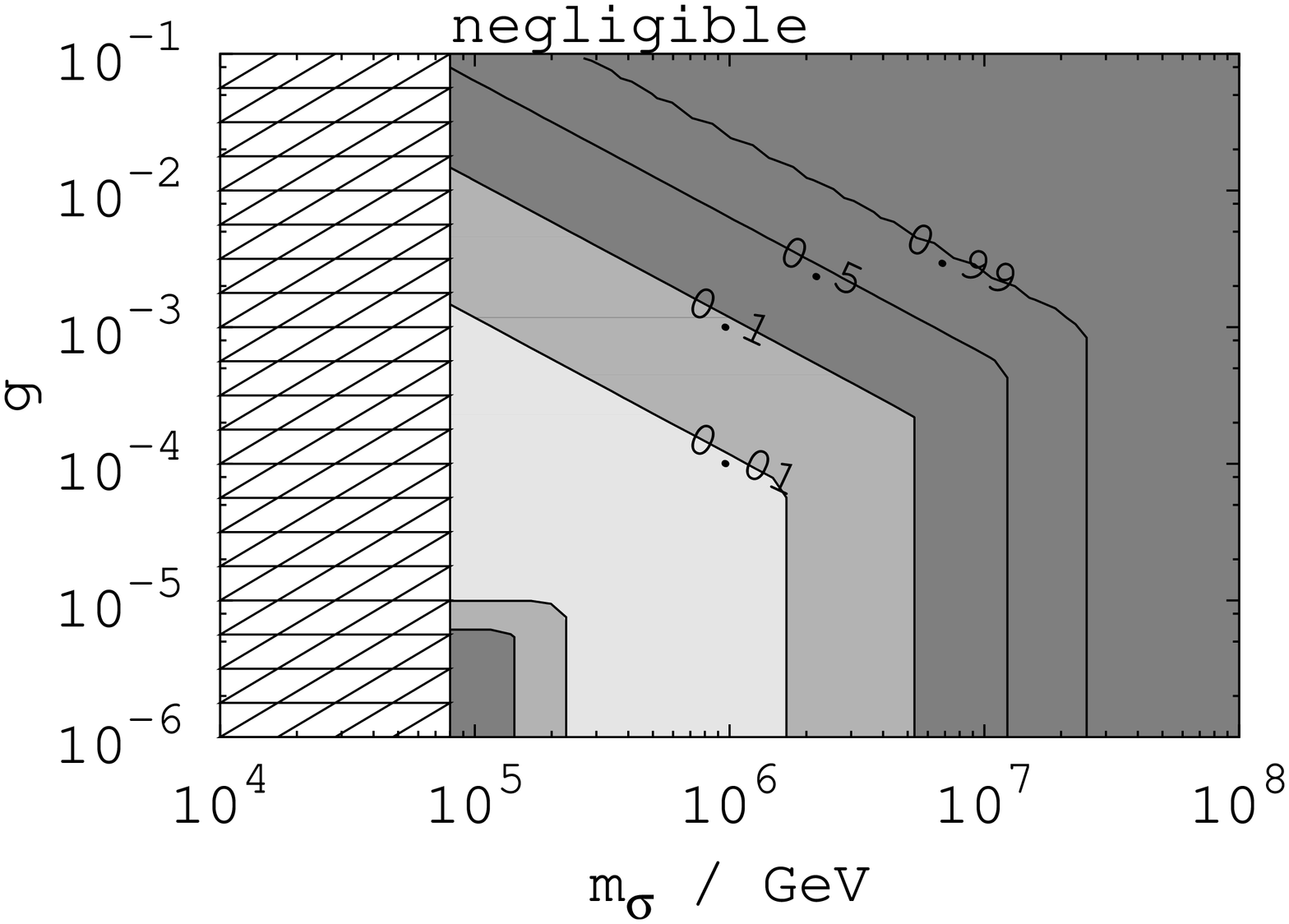}\\
\includegraphics[width=0.45\textwidth, angle=270]{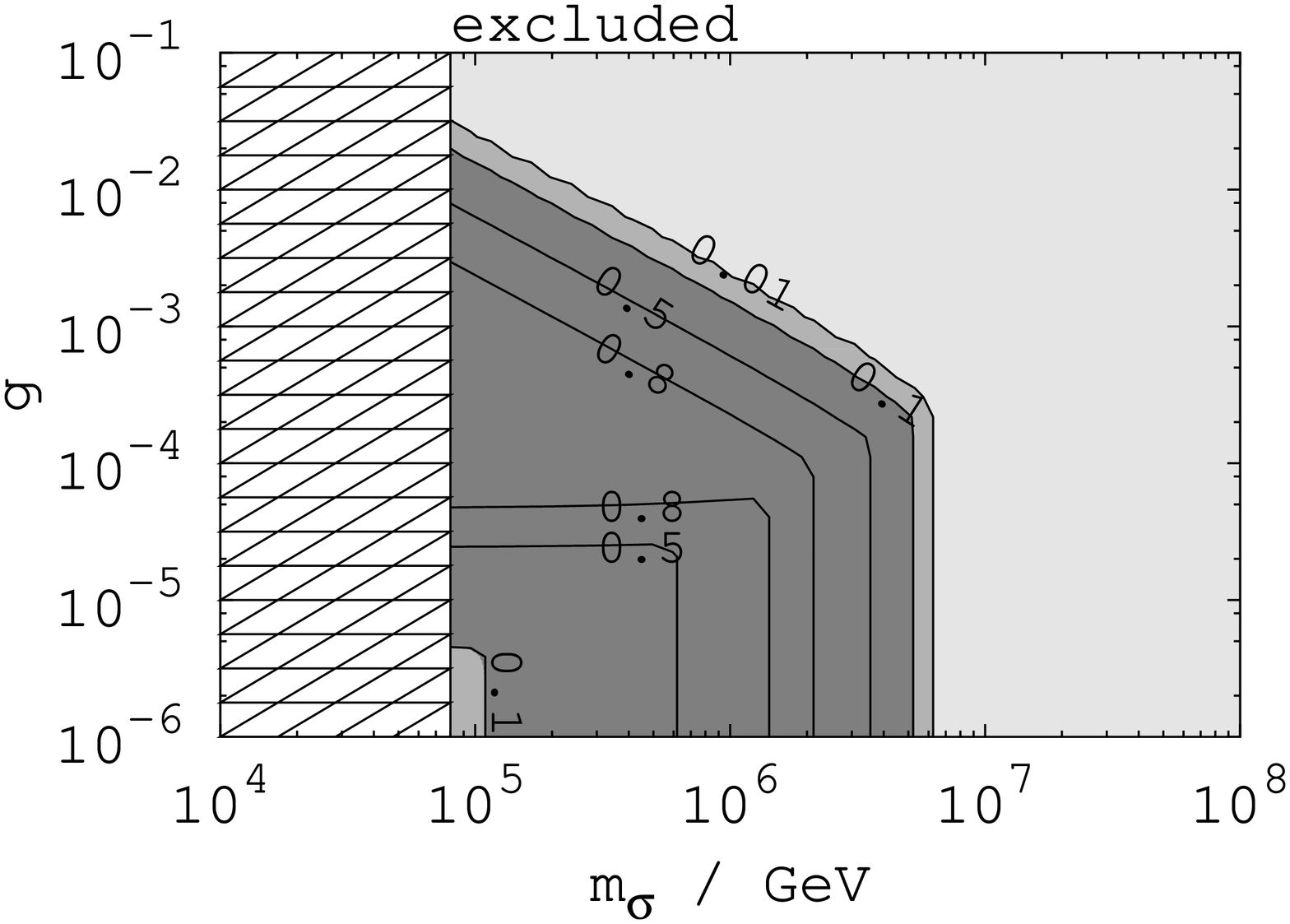}
\caption{\label{fig:higgscont}
$N\to \infty$: probability of obtaining \obs,\ \negl\ and \excl\ as defined in the text. Dark grey has high probability $>10\%$; mid-grey has probability  $>1\%$; light grey low probability $<1\%$; hashed out region is excluded by the BBN bound. For $gH_* < m_\sigma$ the results are independent of $g$. Plotted with $H_* = 10^{10}\GeV$.}
\end{center}
\end{figure}

\fig{fig:higgscont} shows the probability of \obs,\ \negl\ and \excl, where the darkest grey areas have probability $>10\%$, the mid-grey regions have probability $>1\%$ and the lightest grey regions have probability $< 1\%$. Contour lines at 0.5 and 0.8 (0.99 for \negl)\ are also displayed. Low $m_\sigma$ are excluded by the BBN bound (hashed). Recalling that the effective mass can either be dominated by $m_\sigma$ or by $gh_*$ explains the diagonal line, below which the results are independent of $g$.

There are two regions where \obs\ has high probability. The smaller region at low $m_\sigma$, low $g$ occurs when $\rdec \simeq 1$. It only exists for small $\meff$ where the effective decay rate is small and thus the curvaton decays late. It is partially ruled out by the BBN bound $m_\sigma\geq 8\times 10^4\GeV$. It would be completely ruled out by the WIMP bound ($m_\sigma \geq 2\times 10^7\GeV$, see \fig{fig:baremass}), if dark matter was found to consist of thermal relic WIMPs and thus if this bound were applicable. No equivalent bound exists for non-thermal types of dark matter. The larger region extends to all $g$ and is at approximately $m_\sigma = 7 \times 10^6\GeV$ for small $g$, and at smaller $m_\sigma$ for larger $g$. In this large part of the \obs\ region, the curvaton has $\rdec < 1$, meaning that large non-Gaussianity could be possible. The \excl\ region lies in between the two \obs\ regions, and typically produces too high $\zeta$. In the remainder of the parameter space, particularly at large $m_\sigma$, \negl\ dominates and $\zeta$ is typically much lower than $\zobs$. Note that the contours steeply drop off at large $m_\sigma$.

Thus if $N\to \infty$, and if a stable scalar exists with $m_\sigma \simeq 10^7 \GeV$ and quartic self coupling $\lambda_\sigma = 0$, coupled only to the standard model higgs with $0 \leq g < 10^{-3}$, then it is ``probable'' that it would make an observable contribution to the curvature perturbation and also lie within the limits on $\fnl$. Considering now the $\rdec\simeq 1$ curvaton region, if $N\to \infty$, and if a stable scalar exists with $m_\sigma \simeq 10^5 \GeV$, coupled only to the standard model higgs with coupling $0 \leq g < 10^{-4}$, then it is ``probable'' that it could produce the observed curvature perturbation. Another important conclusion is that a scalar with either $m_\sigma \simeq  2\times 10^6\GeV$ and $g \lesssim 10^{-3}$ or smaller $m_\sigma$ and $10^{-4} \lesssim g \lesssim 10^{-3}$ has a high probability of producing $\zeta > \zobs$, which is ruled out by observations. Thus, on probabilistic terms, we could rule out this region of parameter space. Note that these considerations apply to any spectator field fulfilling the above conditions, whether or not it is desired or designed to behave as a curvaton.

For smaller $H_* = 10^9\GeV$, no \excl\ region is seen, and the \obs\ region extends only up to $m_\sigma \simeq 8\times 10^5\GeV$. For $H_* = 10^8\GeV$, the entire parameter space is \negl\ because $\zeta_{max} < 0.1 \zeta_{obs}$.

\subsection{Finite number of $e$-folds}

In the same quadratic potential, we consider the evolution of \obs,\ \negl\ and \excl\ in the case where inflation lasts a finite duration. In this case, the probable regions of the parameter space depend both on $N_{pre}$, the number of $e$-folds of inflation before the last $N_{obs}$ $e$-folds, and on the initial condition of the field $\sigma_0$ before the entire period of inflation. We assume an initial peaked distribution for the curvaton field ($w_0 = 0$), and choose two different values for the central value, $\sigma_0 = 0$ and $\sigma_0 = \mpl$. These are motivated by being the minimum of the curvaton's potential, and by being the only known scale in the theory.

Our philosophy in this paper is not to fine tune the initial $\sigma_0$. If we were to do this, then we could choose $\sigma_0$ and $N_{pre}$ to give observable $\zeta$ for most $m_\sigma$. Although this is a common method of choosing the initial conditions for the curvaton model, it is precisely what we wish to avoid.

The time scales for the evolution of $P(\sigma,N)$ are given by $N_{rel}$ and $N_{dec}$ \eq{rel}. The timescale are longer for smaller $m_\sigma$, but do not depend on $\sigma_0$. However, for $\sigma_0 = \mpl$, the central value of the distribution must move much further to reach the equilibrium distribution, which is centered around zero. Thus, the evolution in that case is expected (and confirmed) to take somewhat longer.\footnote{Note that the approximation for the curvaton's effective mass is not necessarily valid for $\Npre \lesssim 100$ because the higgs needs some time to reach its equilibrium distribution (see \sect{karipaper}).}

\begin{figure}
\begin{center}
\includegraphics[ width=0.25\textwidth,angle=270]{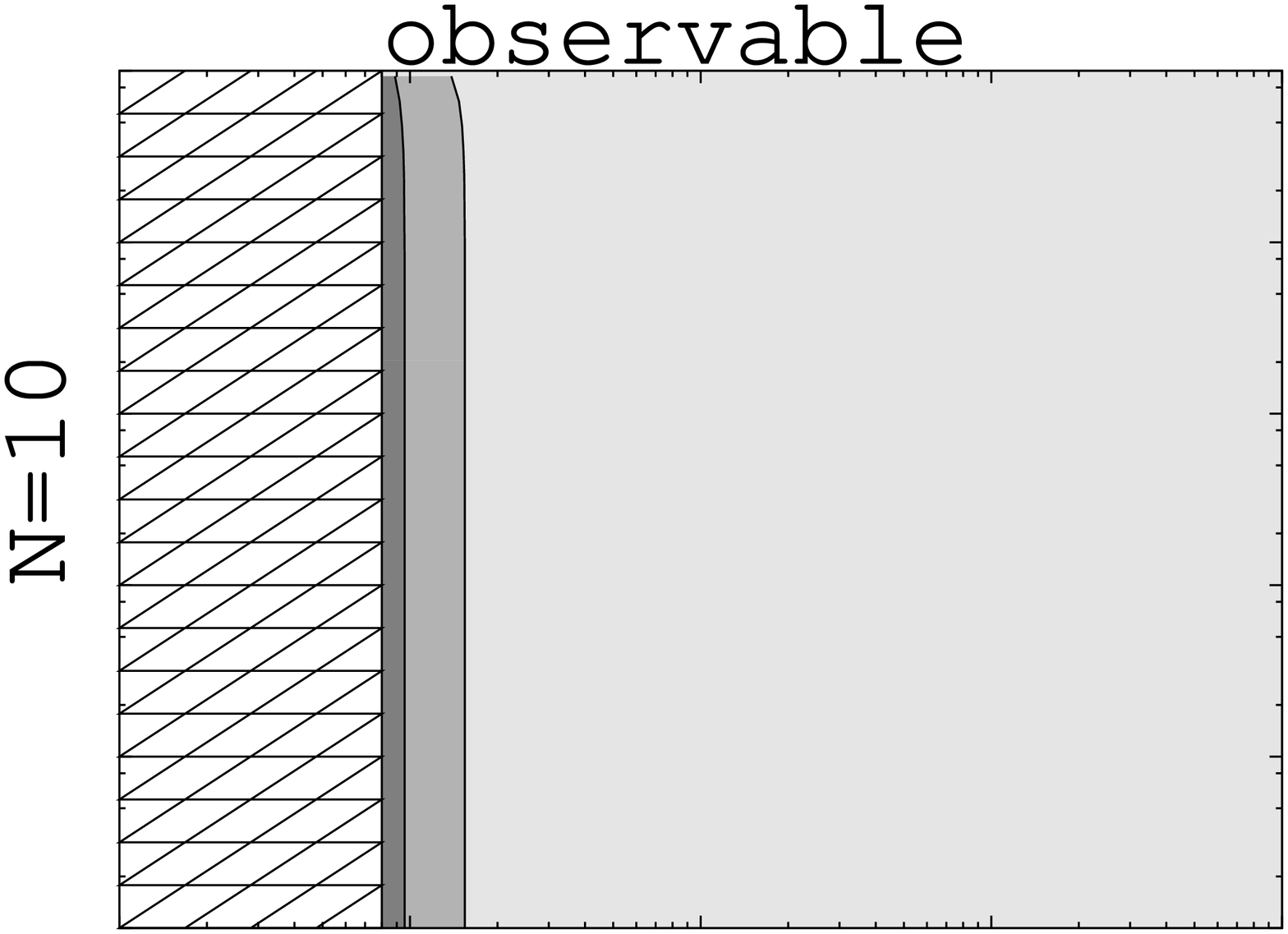} \hspace{-1cm}
\includegraphics[ width=0.25\textwidth,angle=270]{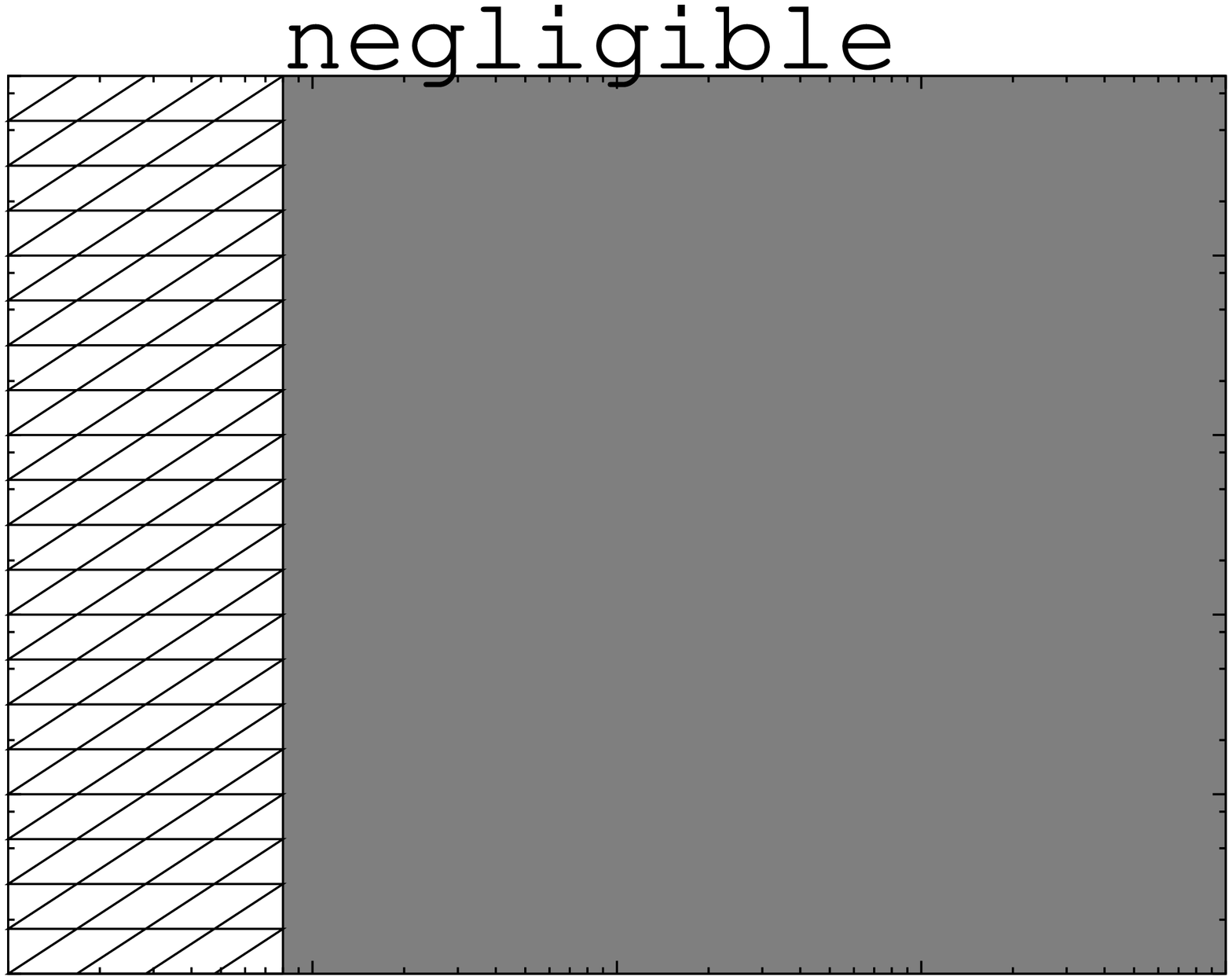}
\hspace{-0.95cm}
\includegraphics[ width=0.25\textwidth,angle=270]{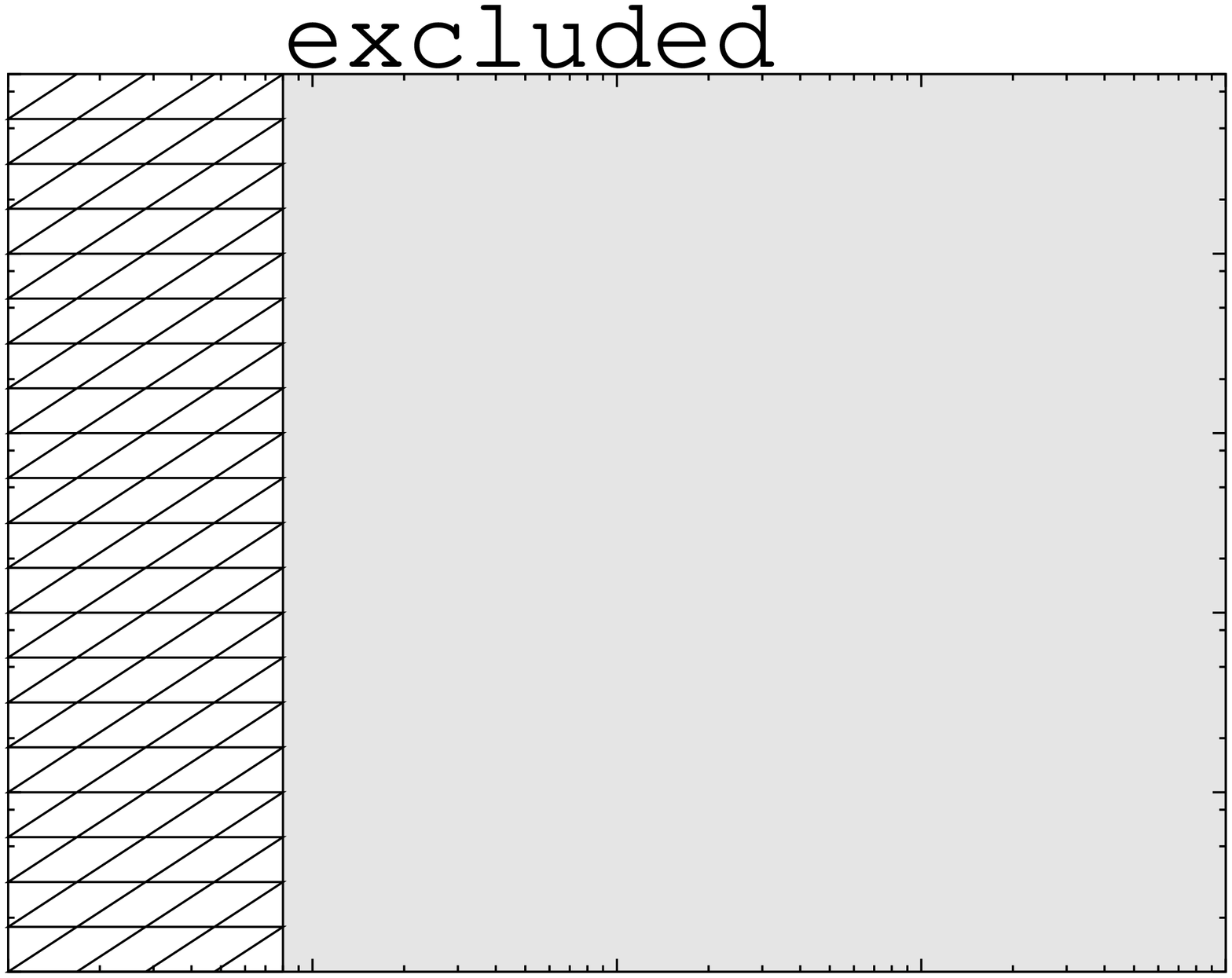} \\
\vspace{-0.3cm}
\includegraphics[ width=0.25\textwidth,angle=270]{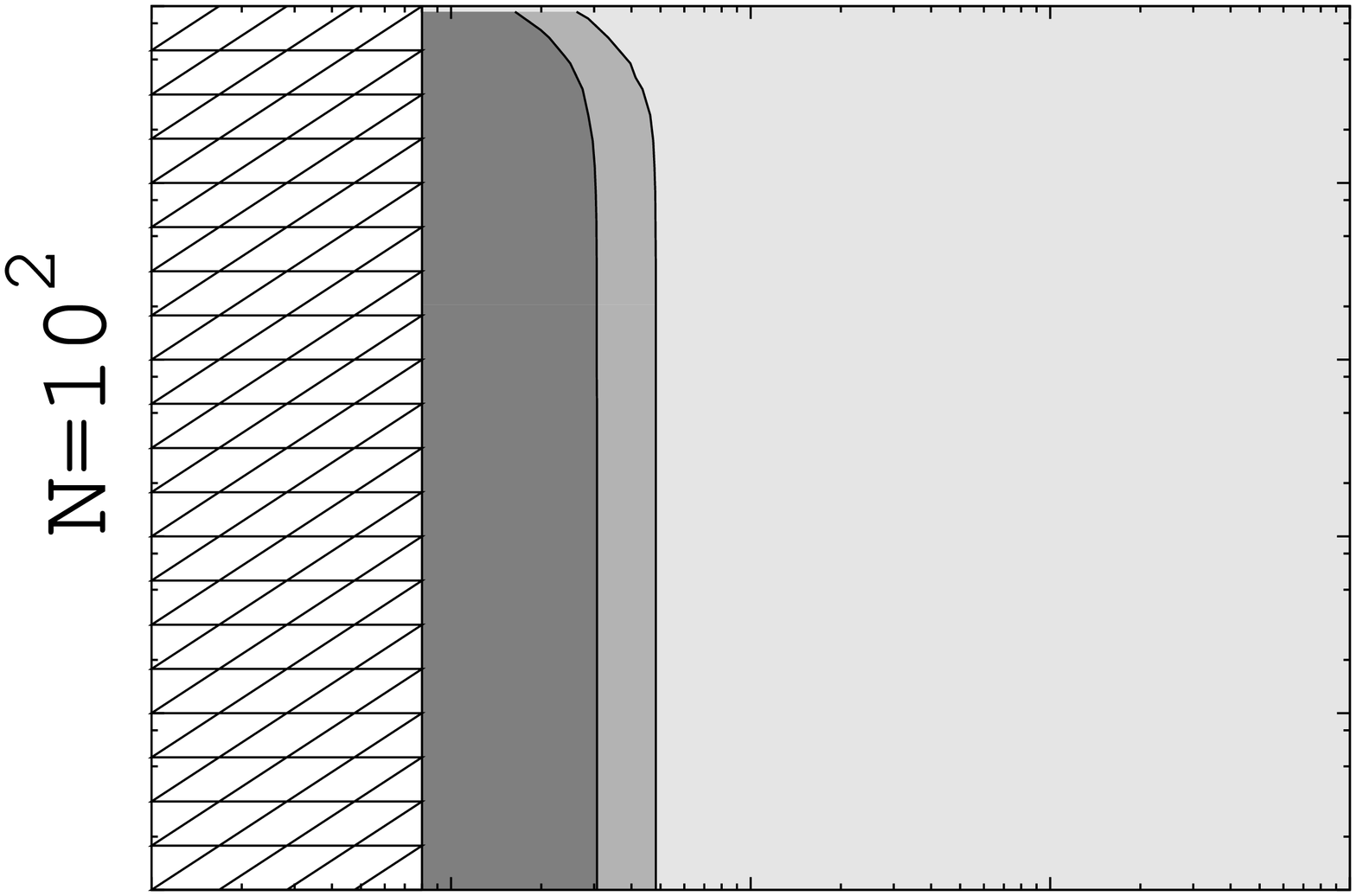} \hspace{-1cm}
\includegraphics[ width=0.25\textwidth,angle=270]{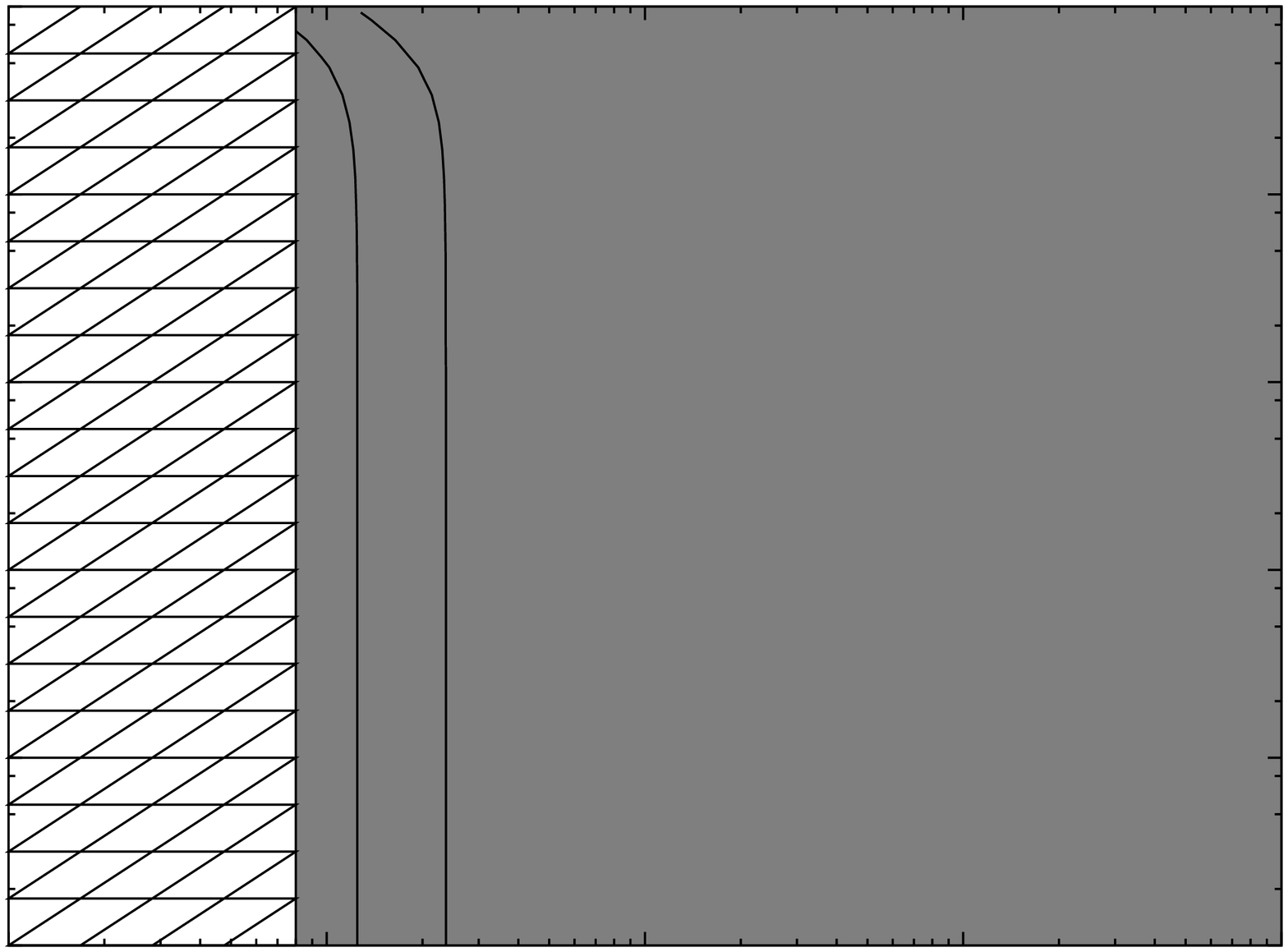}
\hspace{-0.95cm}
\includegraphics[ width=0.25\textwidth,angle=270]{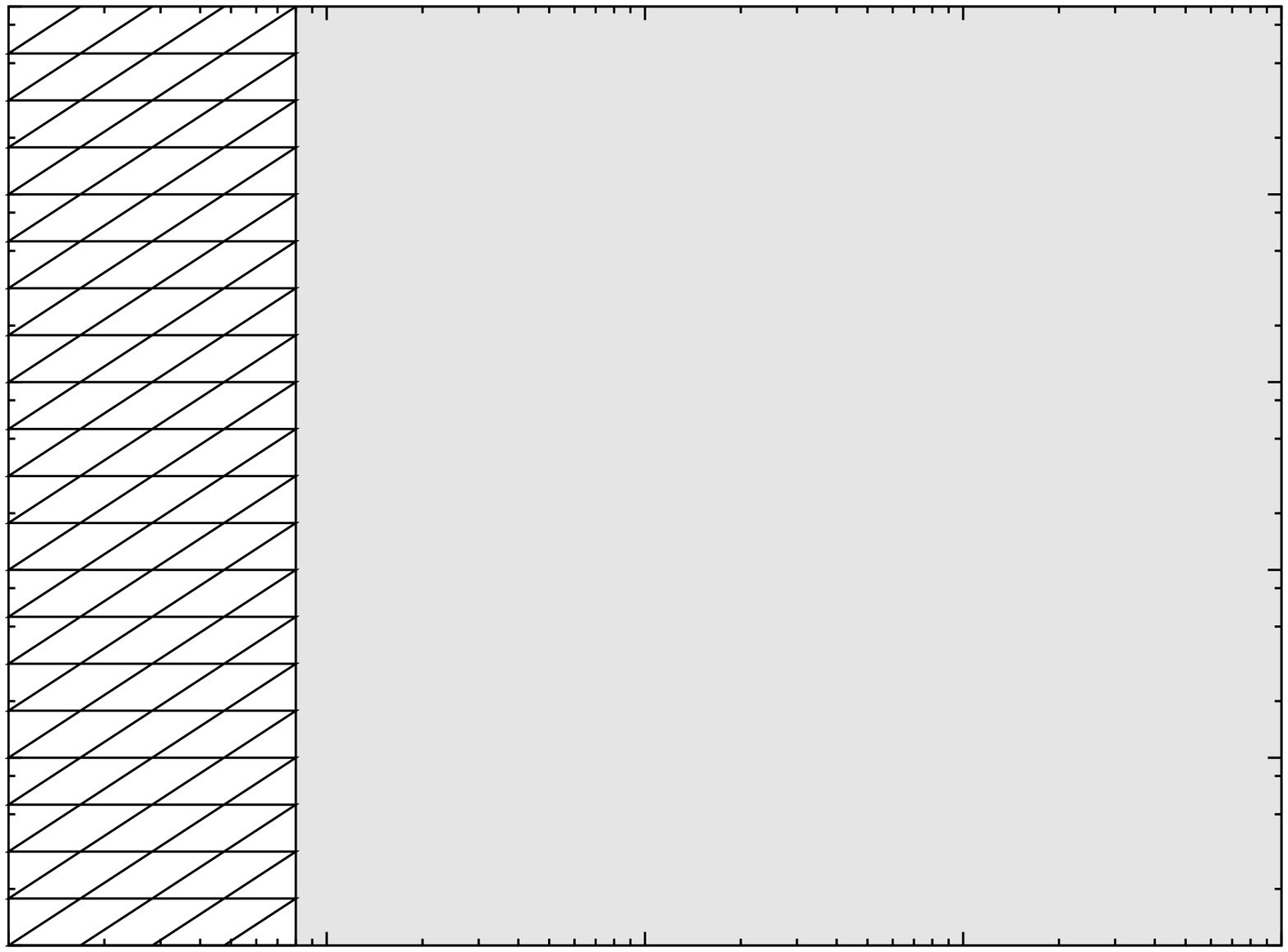} \\
\vspace{-0.3cm}
\includegraphics[ width=0.25\textwidth,angle=270]{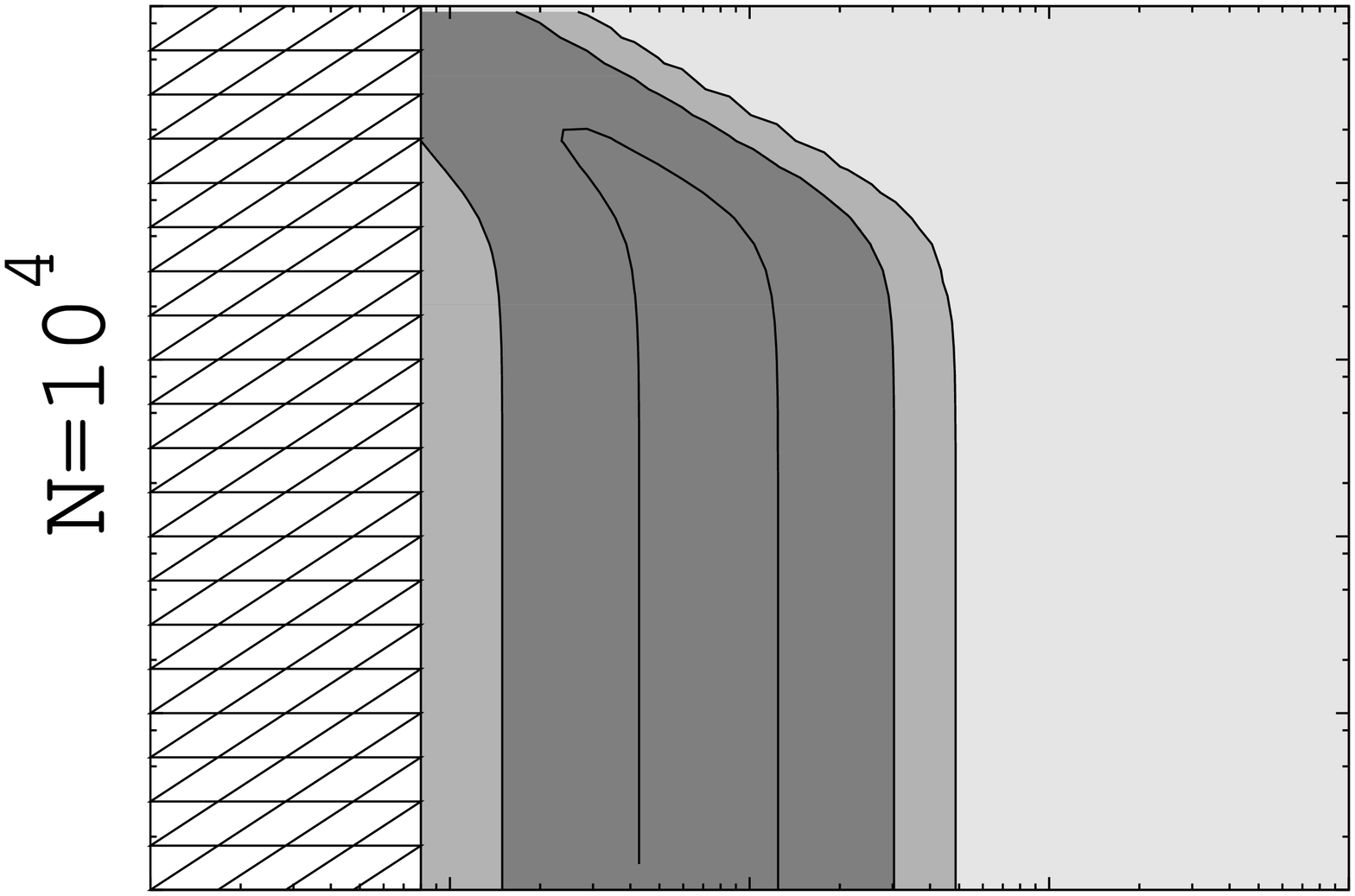} \hspace{-1cm}
\includegraphics[ width=0.25\textwidth,angle=270]{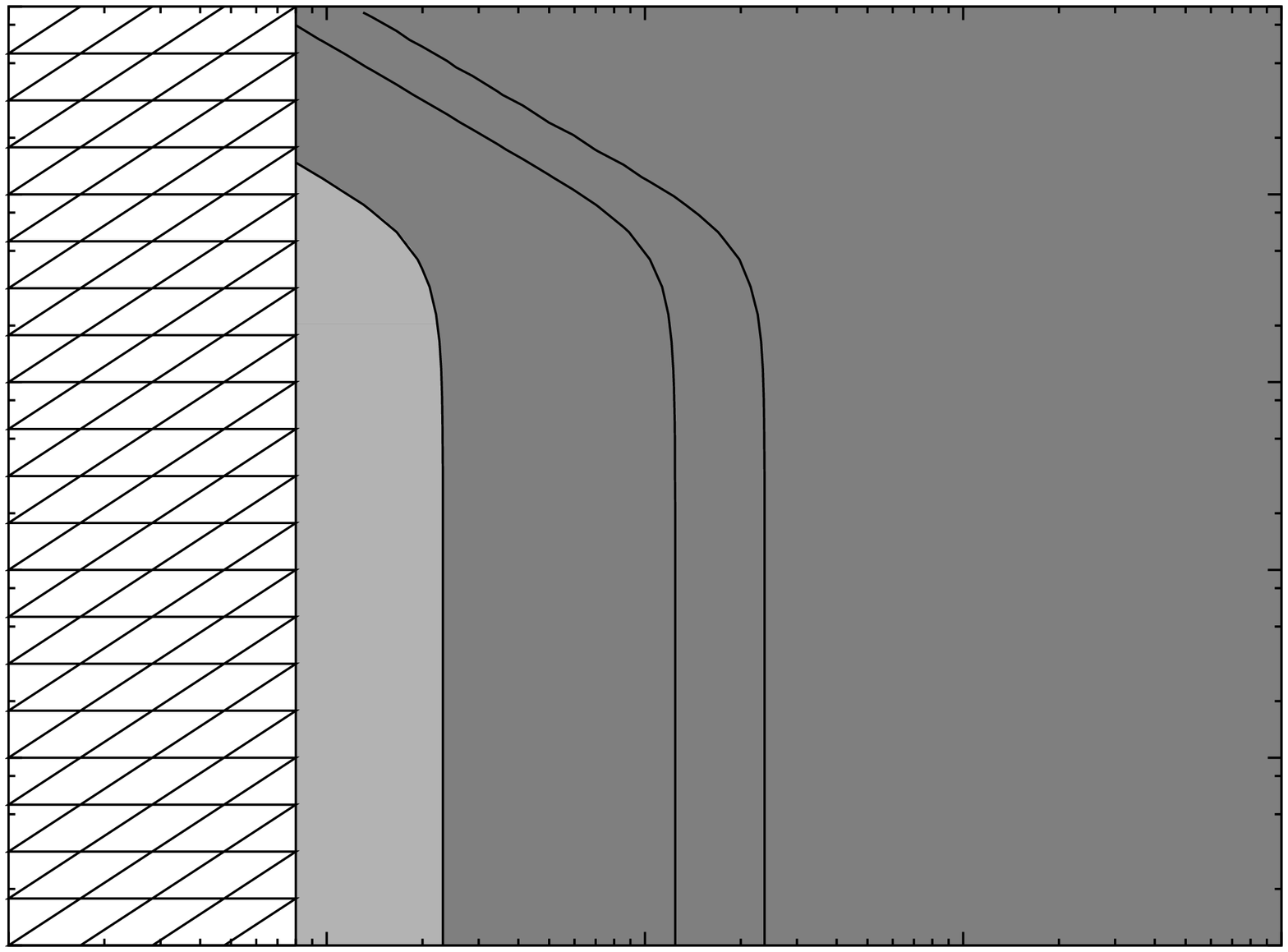}
\hspace{-0.95cm}
\includegraphics[ width=0.25\textwidth,angle=270]{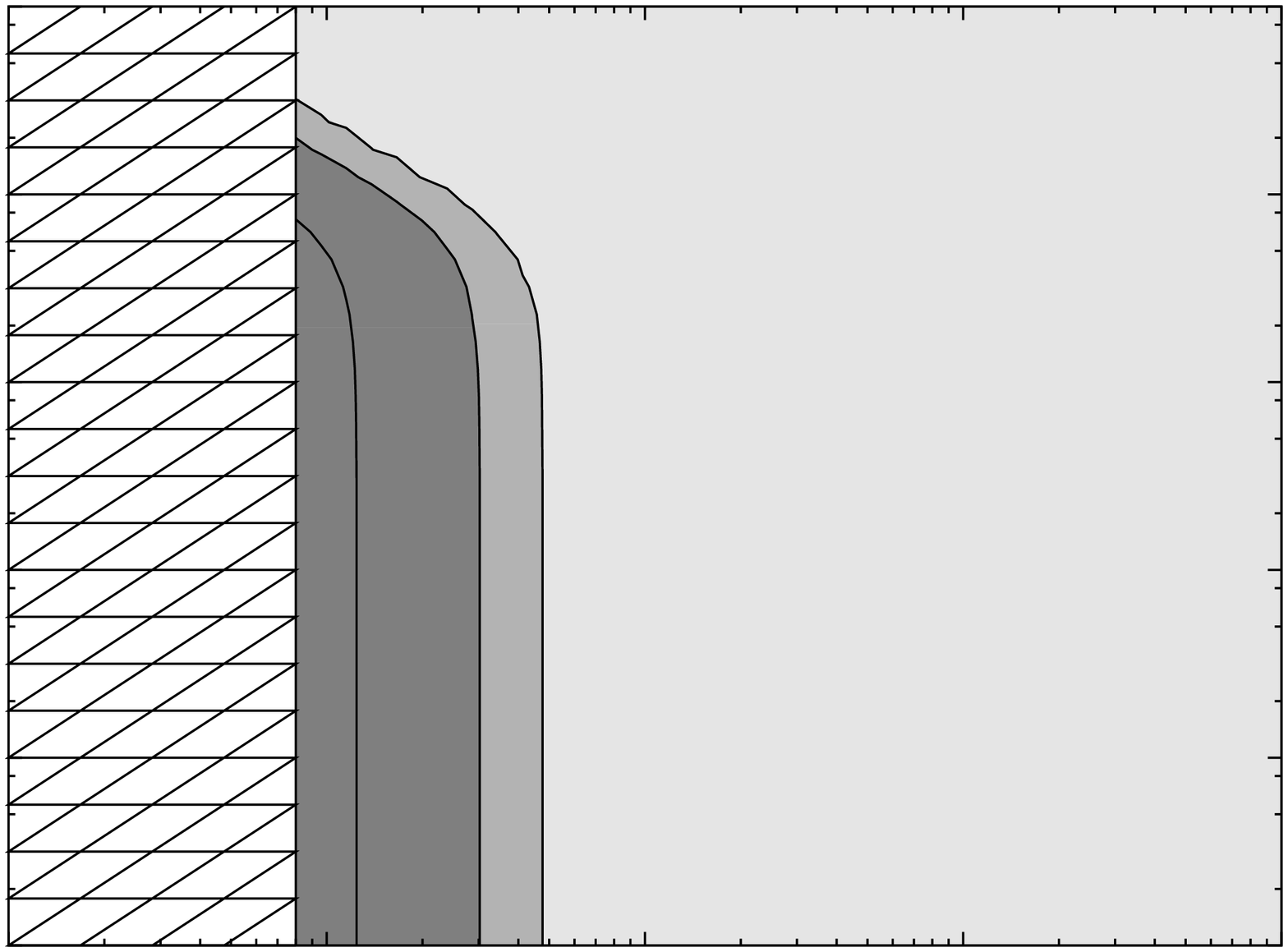} \\
\vspace{-0.3cm}
\includegraphics[ width=0.25\textwidth,angle=270]{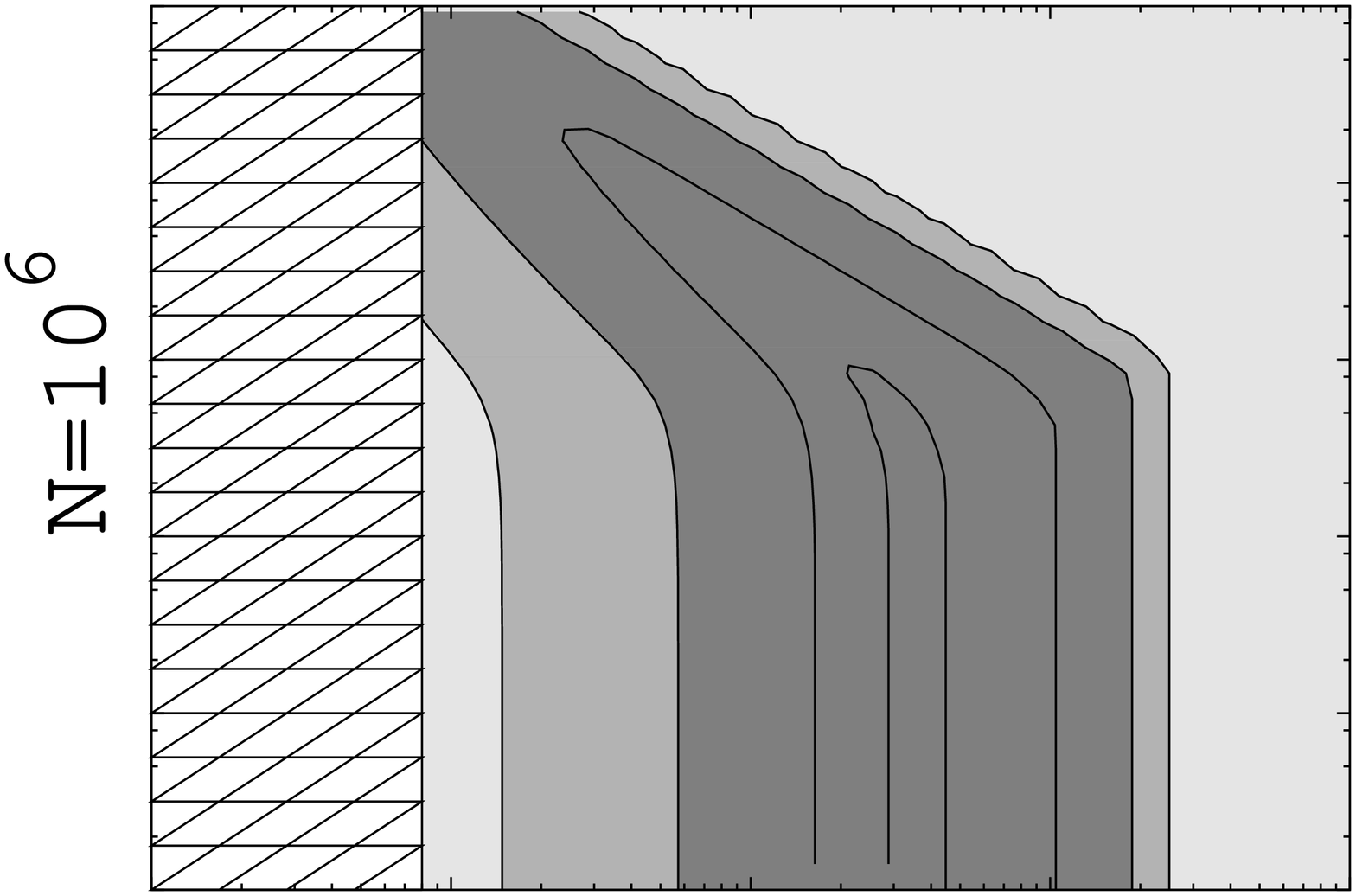} \hspace{-1cm}
\includegraphics[ width=0.25\textwidth,angle=270]{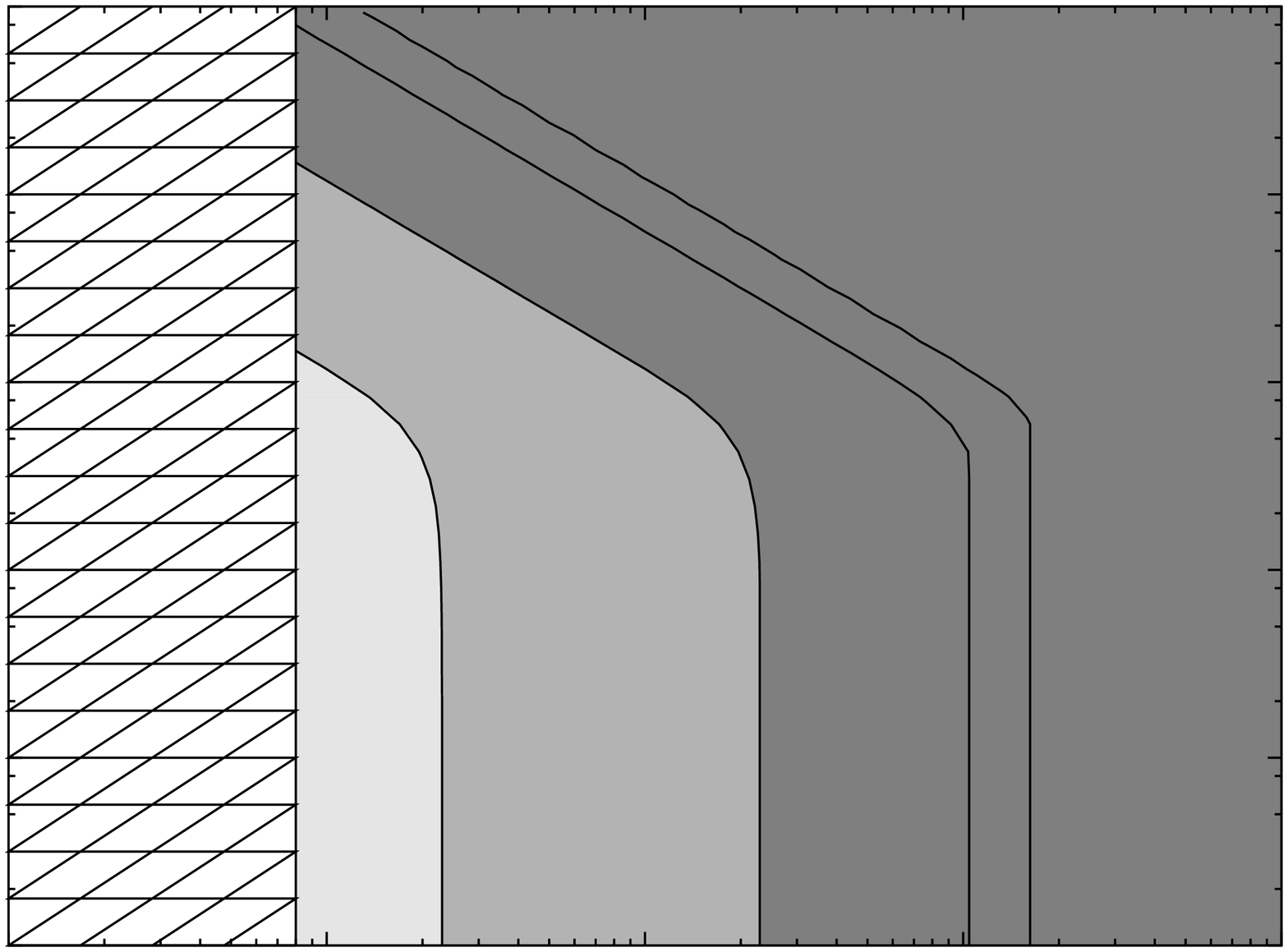}
\hspace{-0.95cm}
\includegraphics[ width=0.25\textwidth,angle=270]{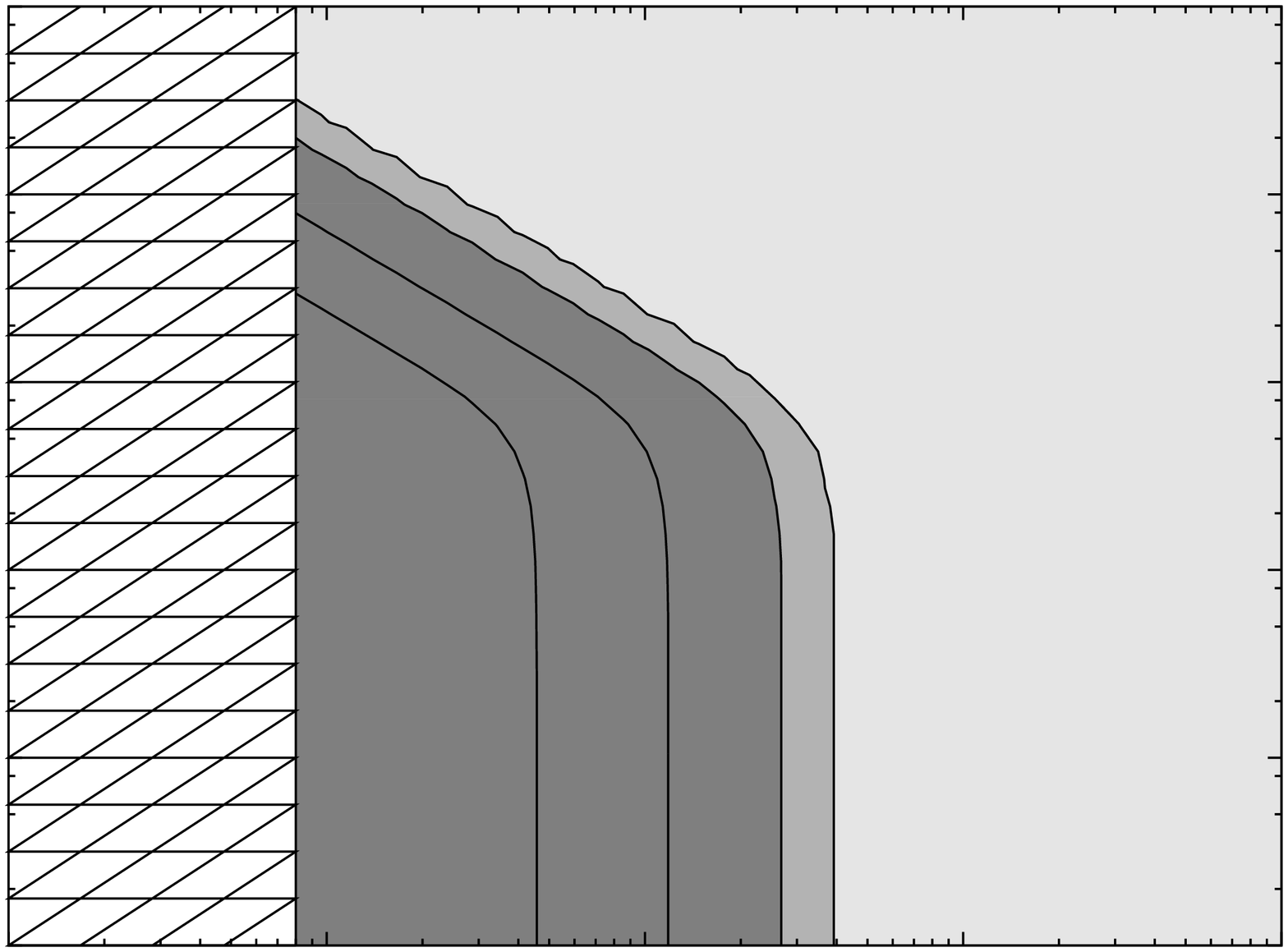} \\
\vspace{-0.3cm}
\includegraphics[ width=0.25\textwidth,angle=270]{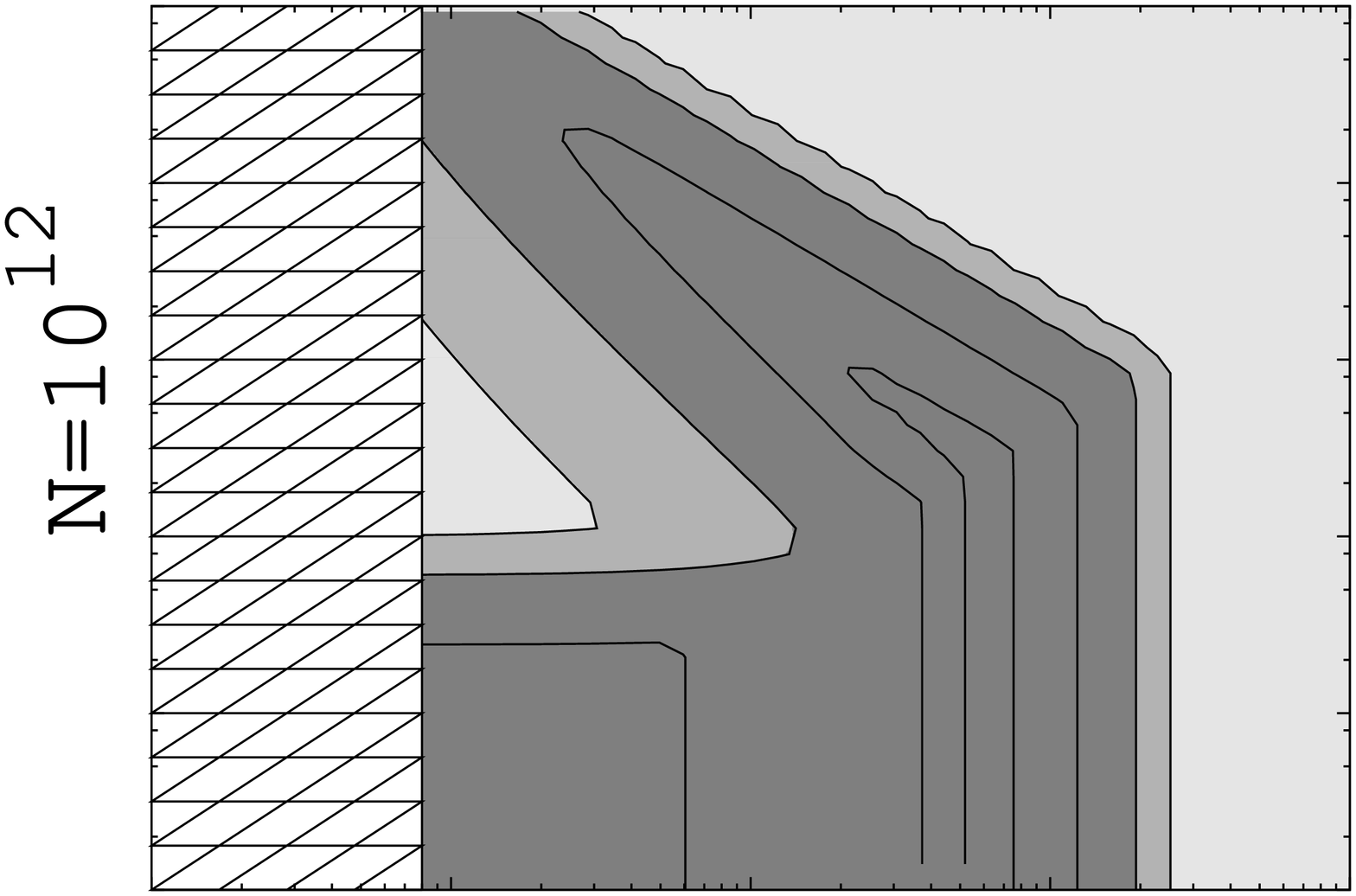} \hspace{-1cm}
\includegraphics[ width=0.25\textwidth,angle=270]{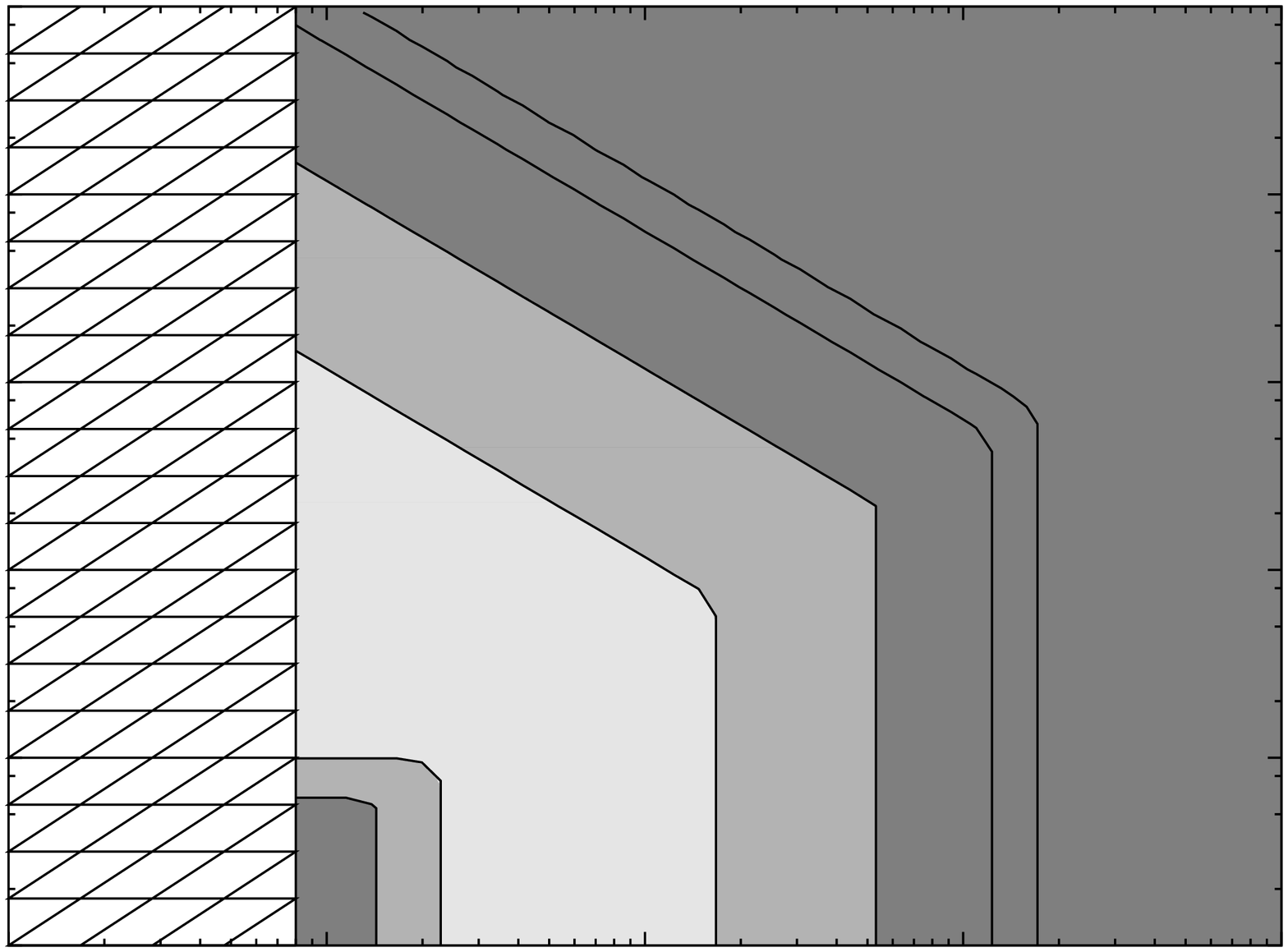}
\hspace{-0.95cm}
\includegraphics[ width=0.25\textwidth,angle=270]{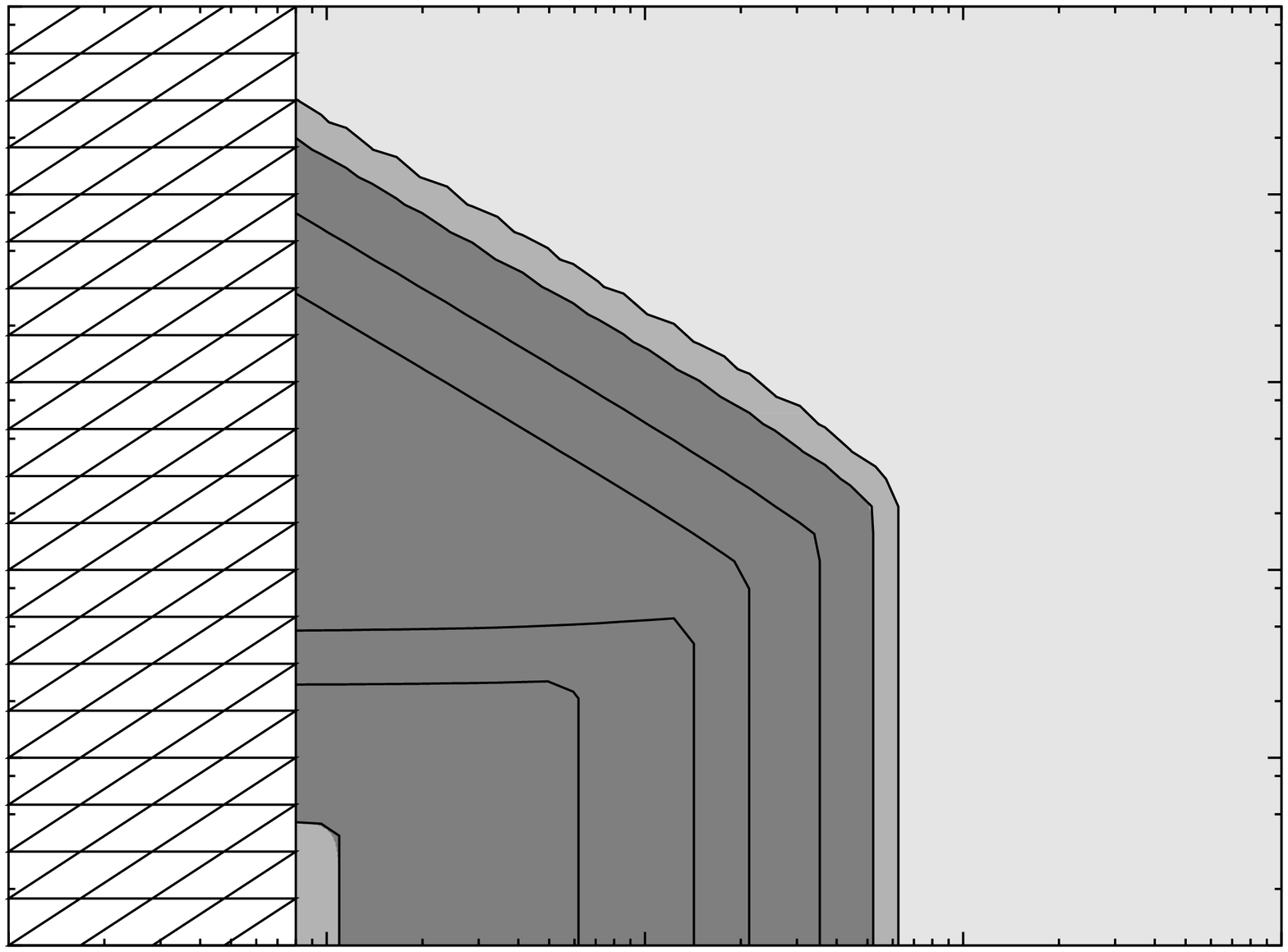} 
\caption{\label{fig:Nsig_0}
$\sigma_0 = 0$: regions \obs,\ \negl\ and \excl\ as defined in the text, for increasing $N_{pre} = 10, 10^2, 10^4, 10^6,10^{12}$. Axes, shading and parameters identical to \fig{fig:higgscont}.}
\end{center}
\vspace{-0.3cm}
\end{figure}

\begin{figure}
\begin{center}
\includegraphics[ width=0.25\textwidth,angle=270]{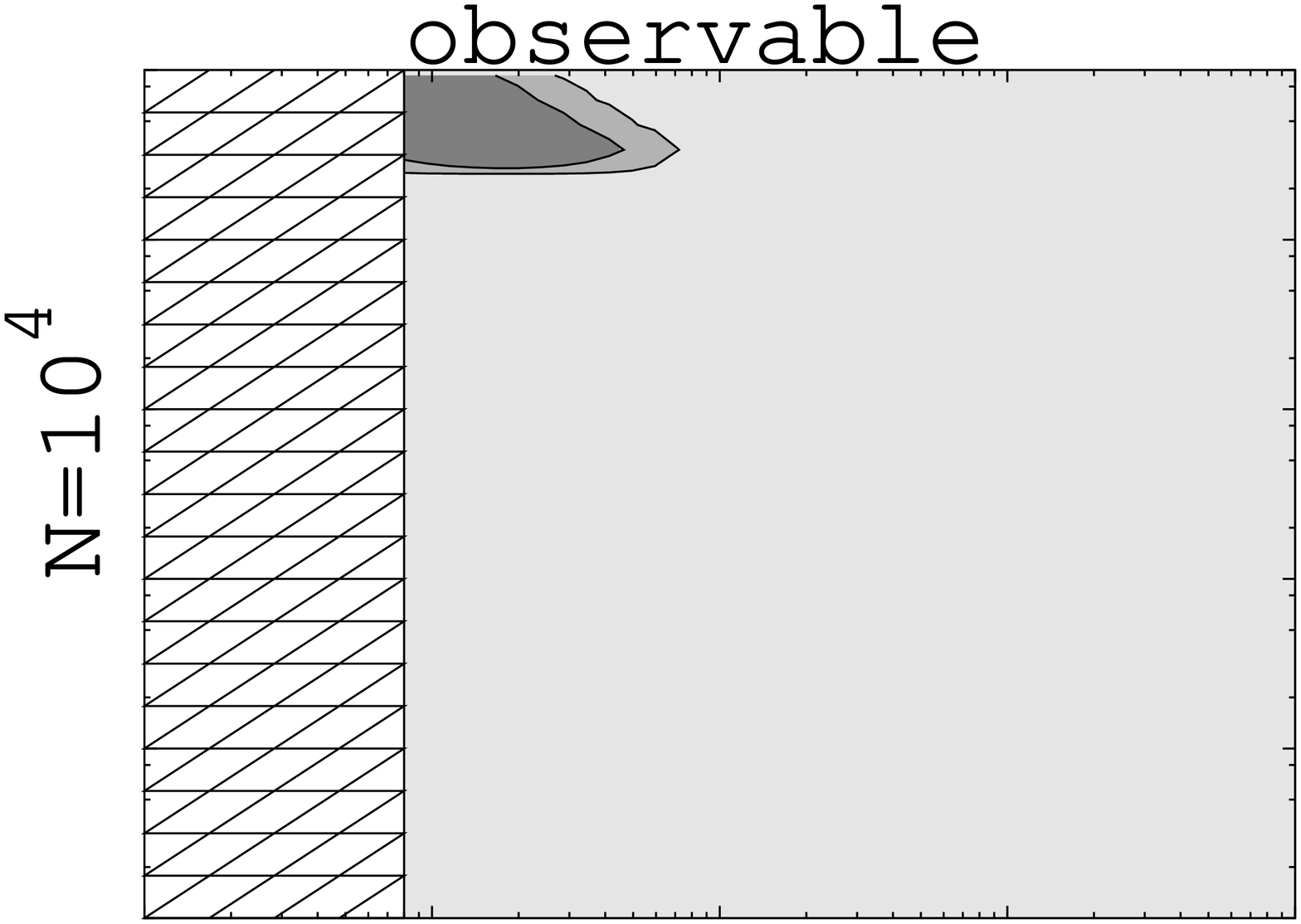} \hspace{-1cm}
\includegraphics[ width=0.25\textwidth,angle=270]{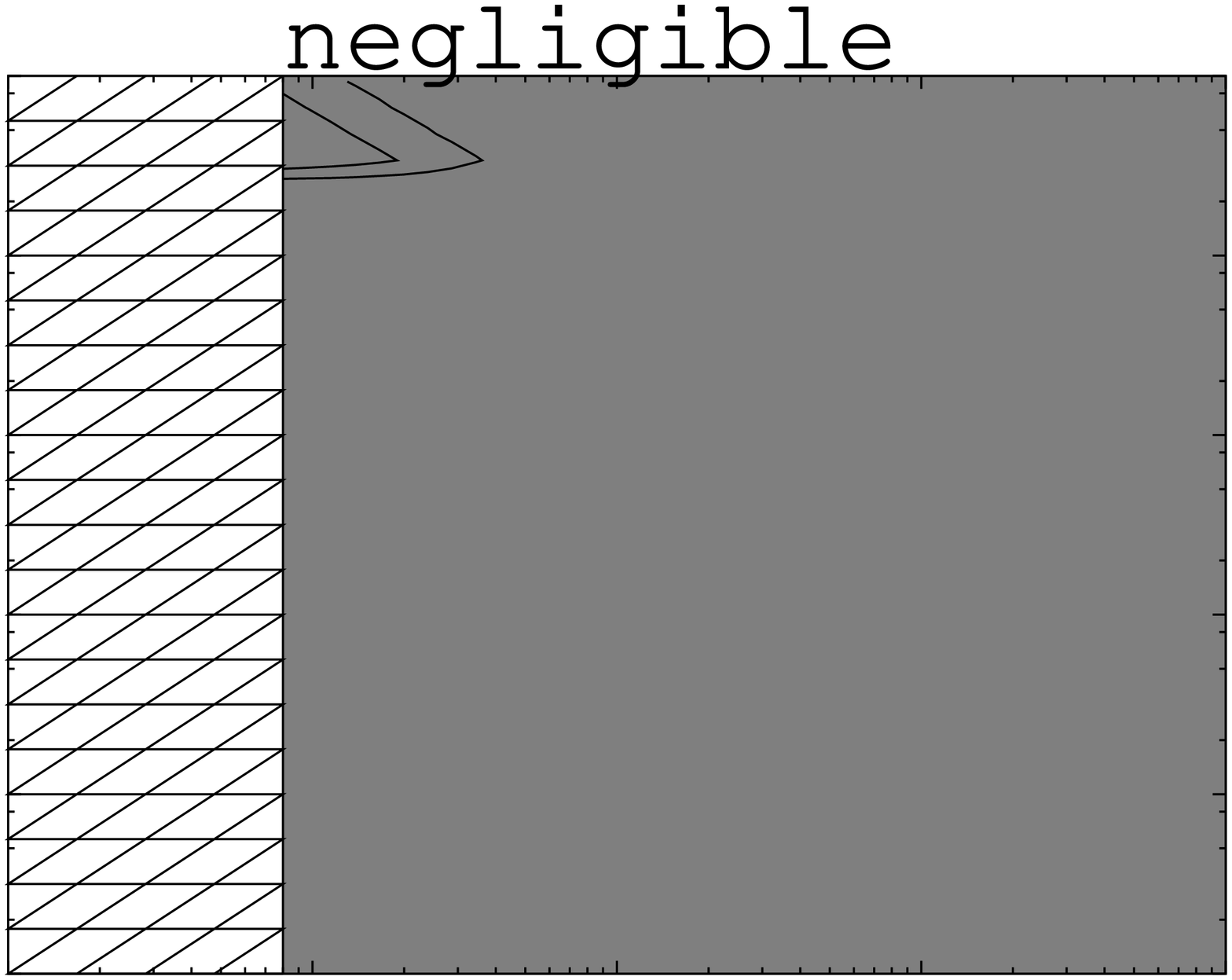}
\hspace{-0.95cm}
\includegraphics[ width=0.25\textwidth,angle=270]{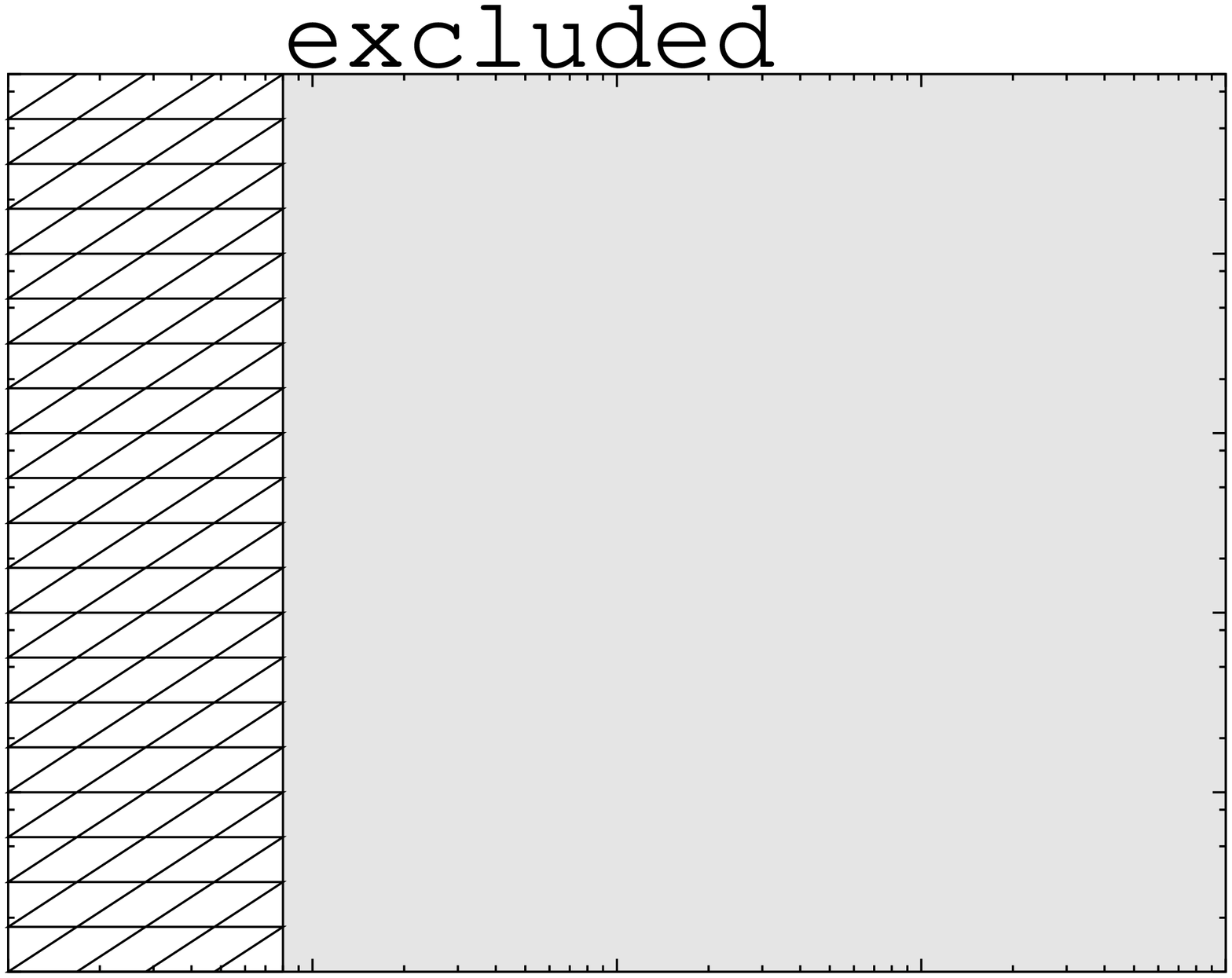} \\
\vspace{-0.3cm}
\includegraphics[ width=0.25\textwidth,angle=270]{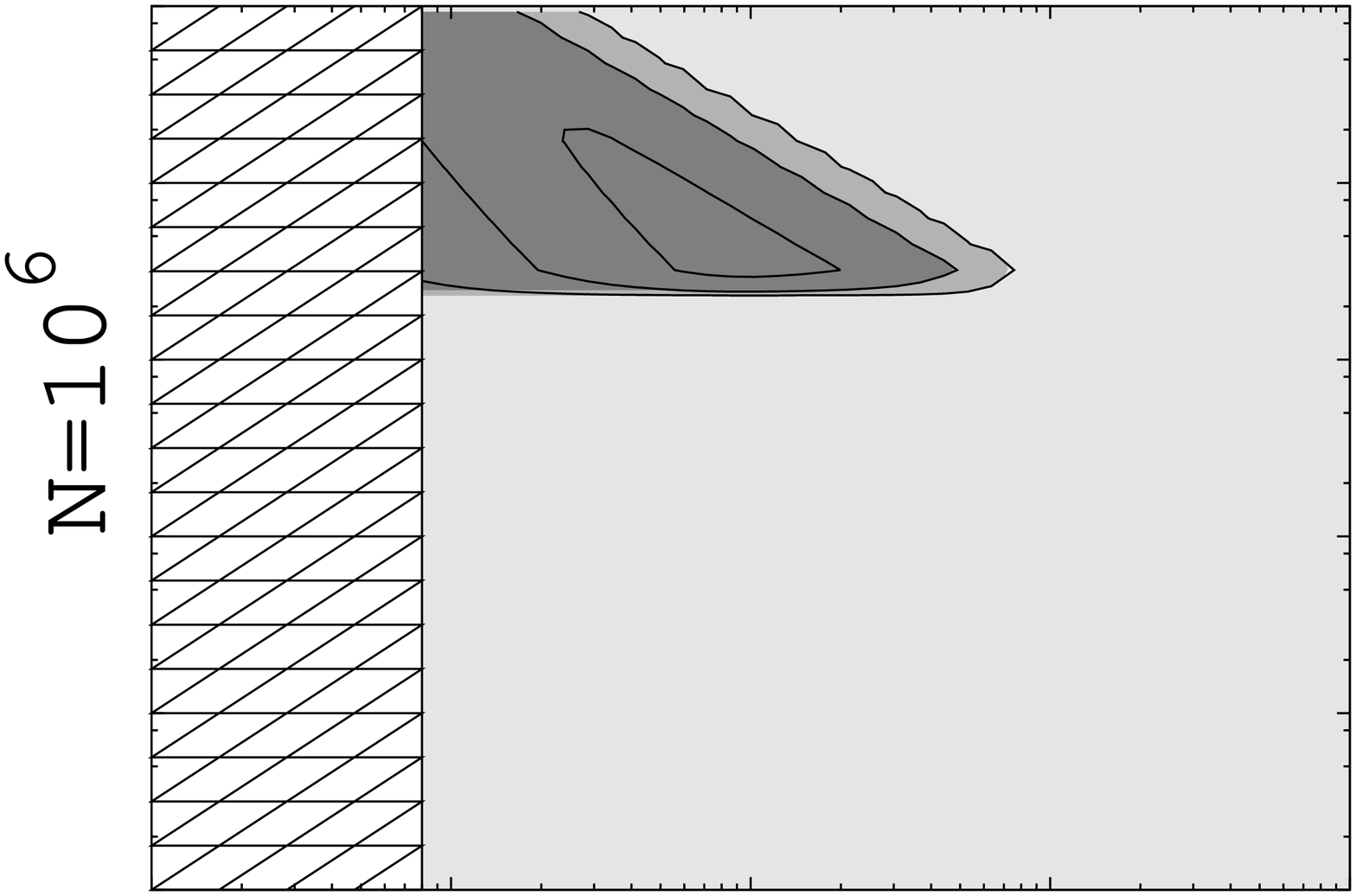} \hspace{-1cm}
\includegraphics[ width=0.25\textwidth,angle=270]{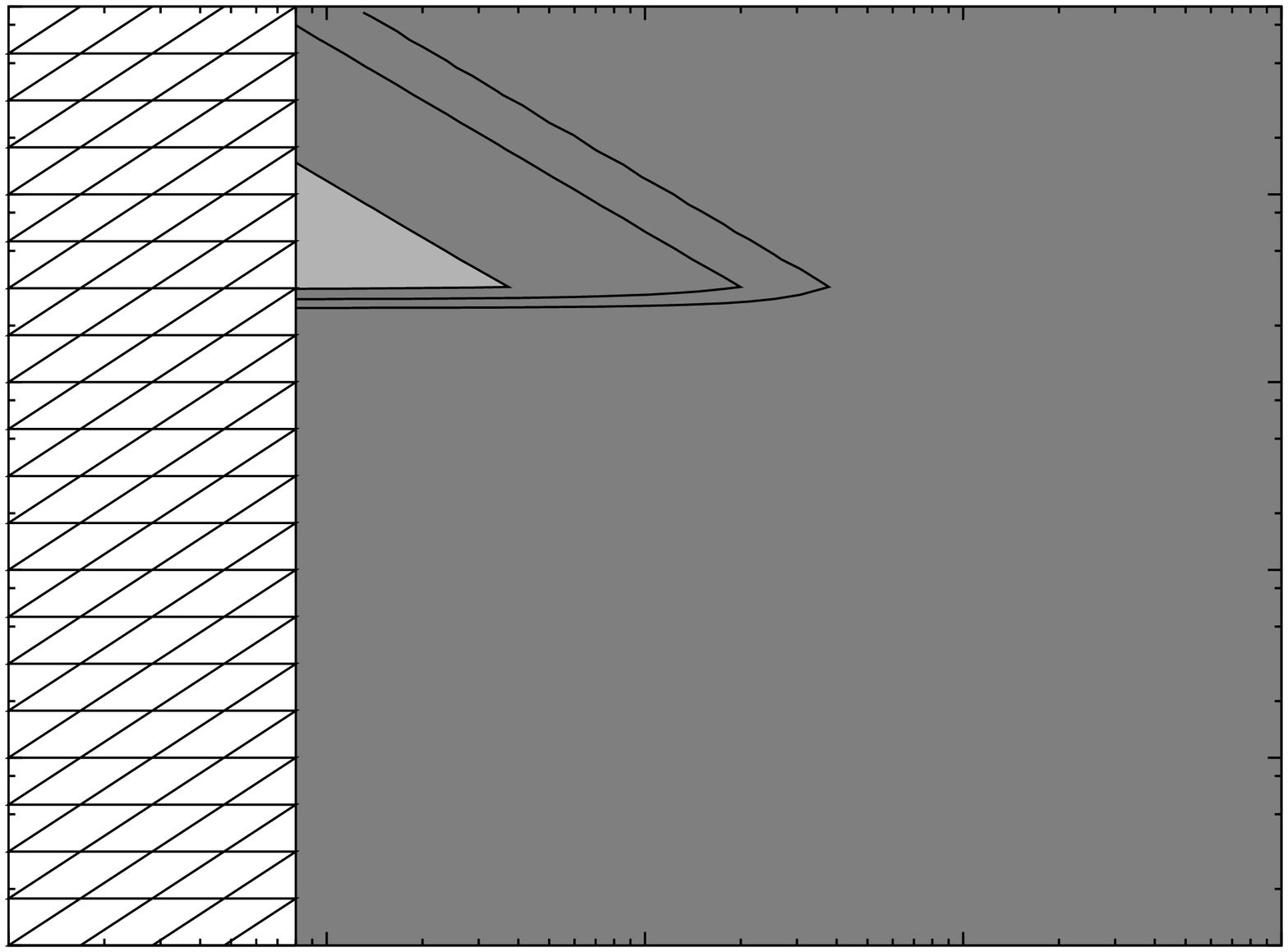}
\hspace{-0.95cm}
\includegraphics[ width=0.25\textwidth,angle=270]{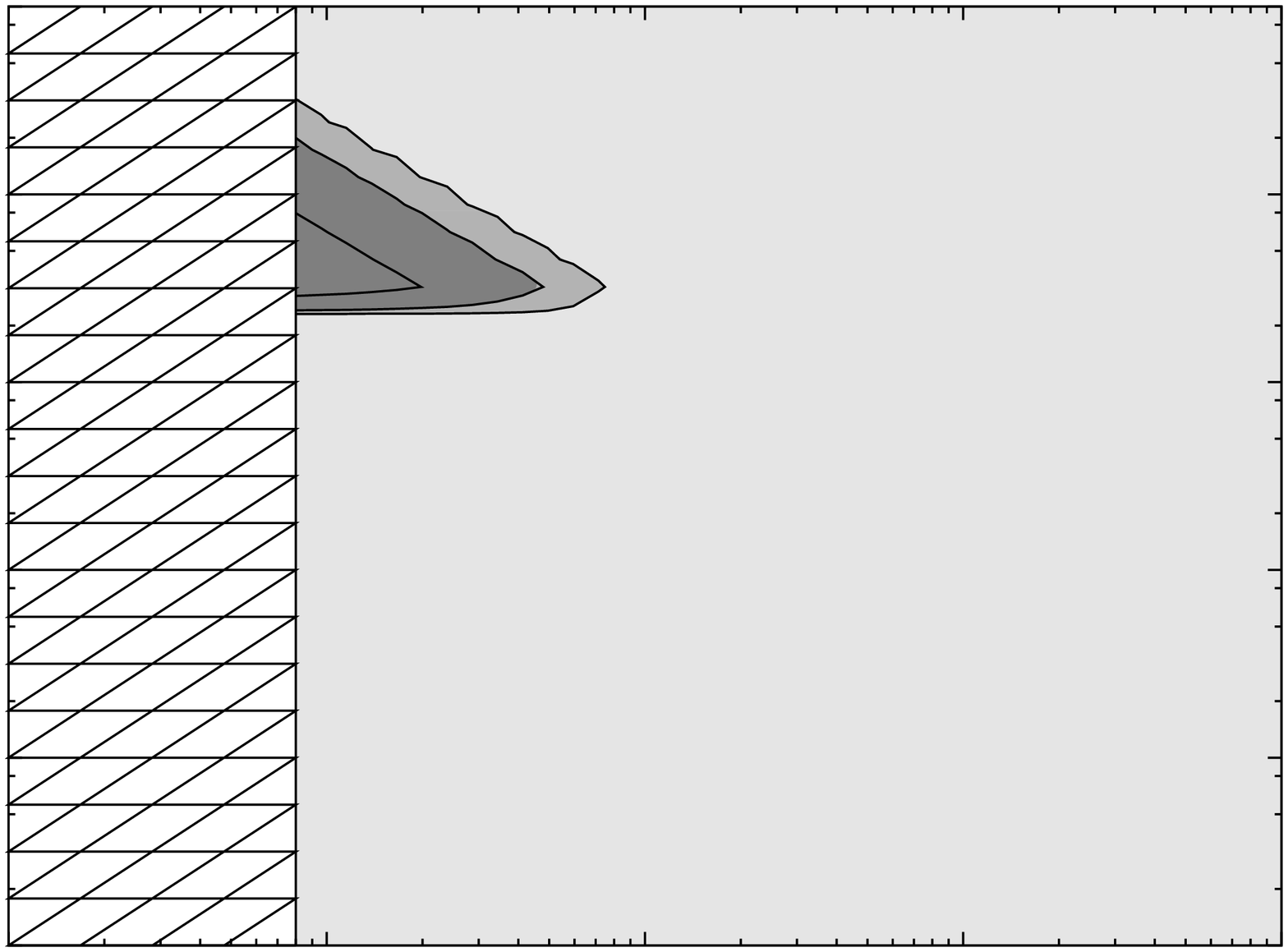} \\
\vspace{-0.3cm}
\includegraphics[ width=0.25\textwidth,angle=270]{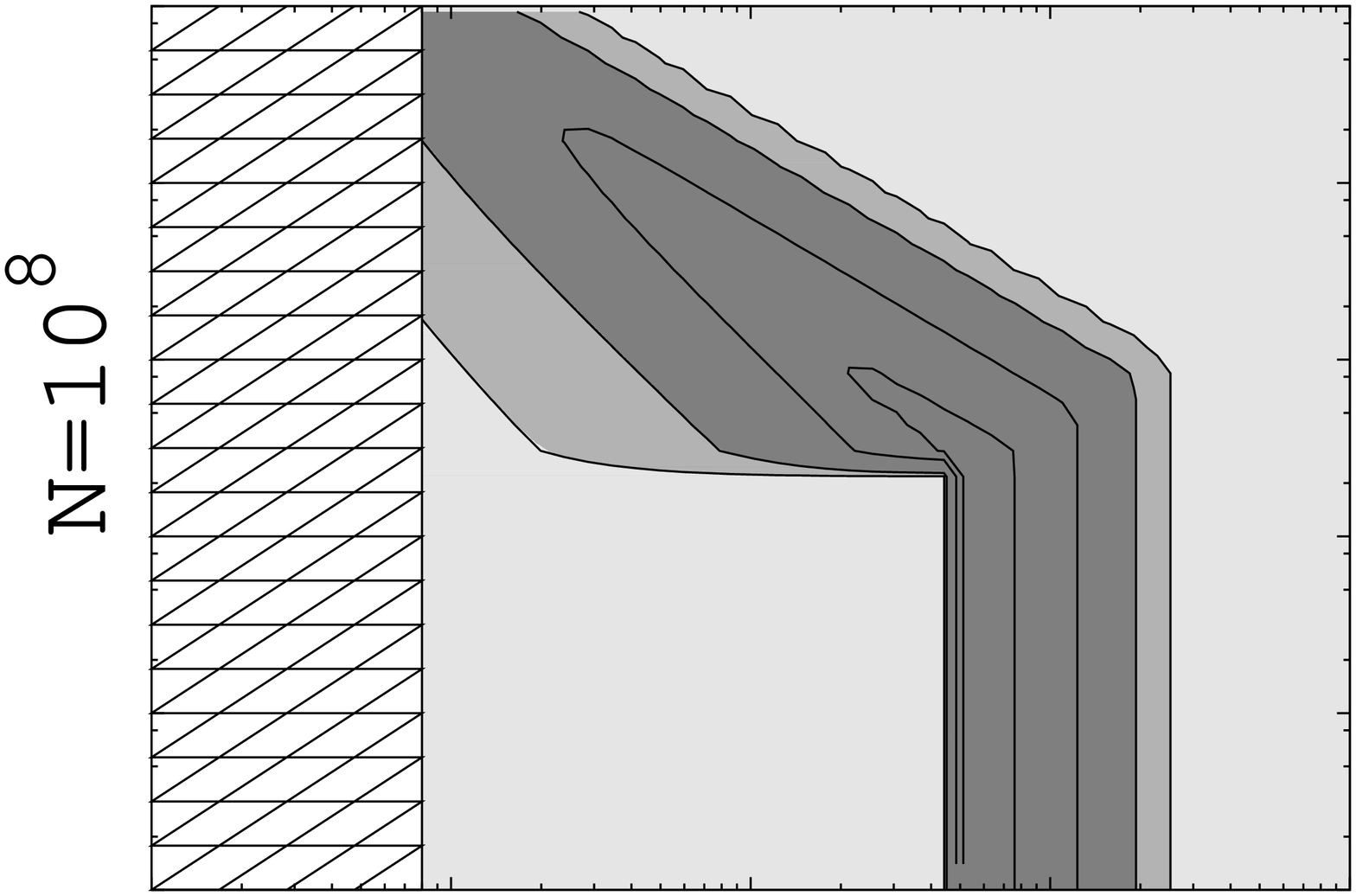} \hspace{-1cm}
\includegraphics[ width=0.25\textwidth,angle=270]{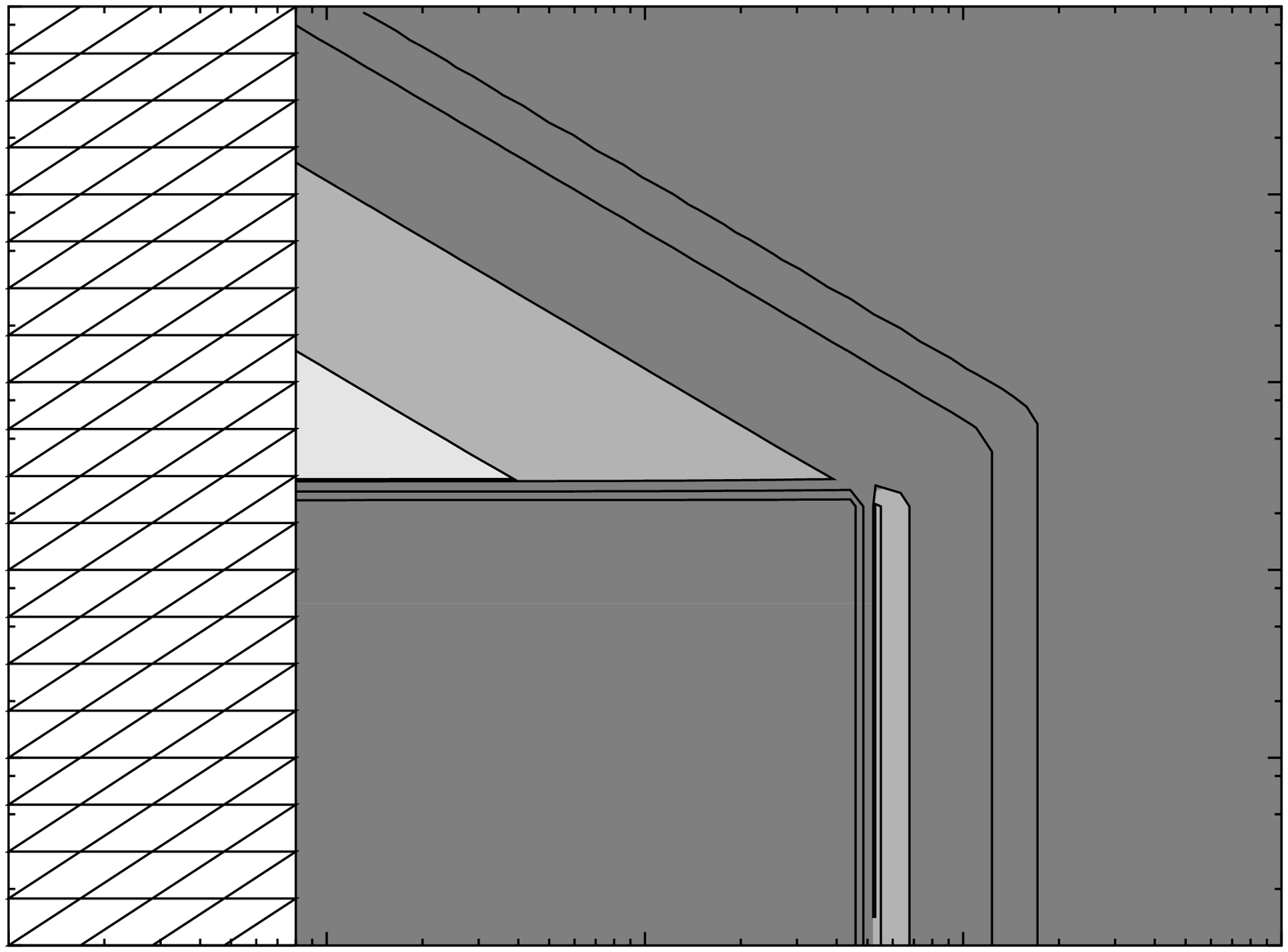}
\hspace{-0.95cm}
\includegraphics[ width=0.25\textwidth,angle=270]{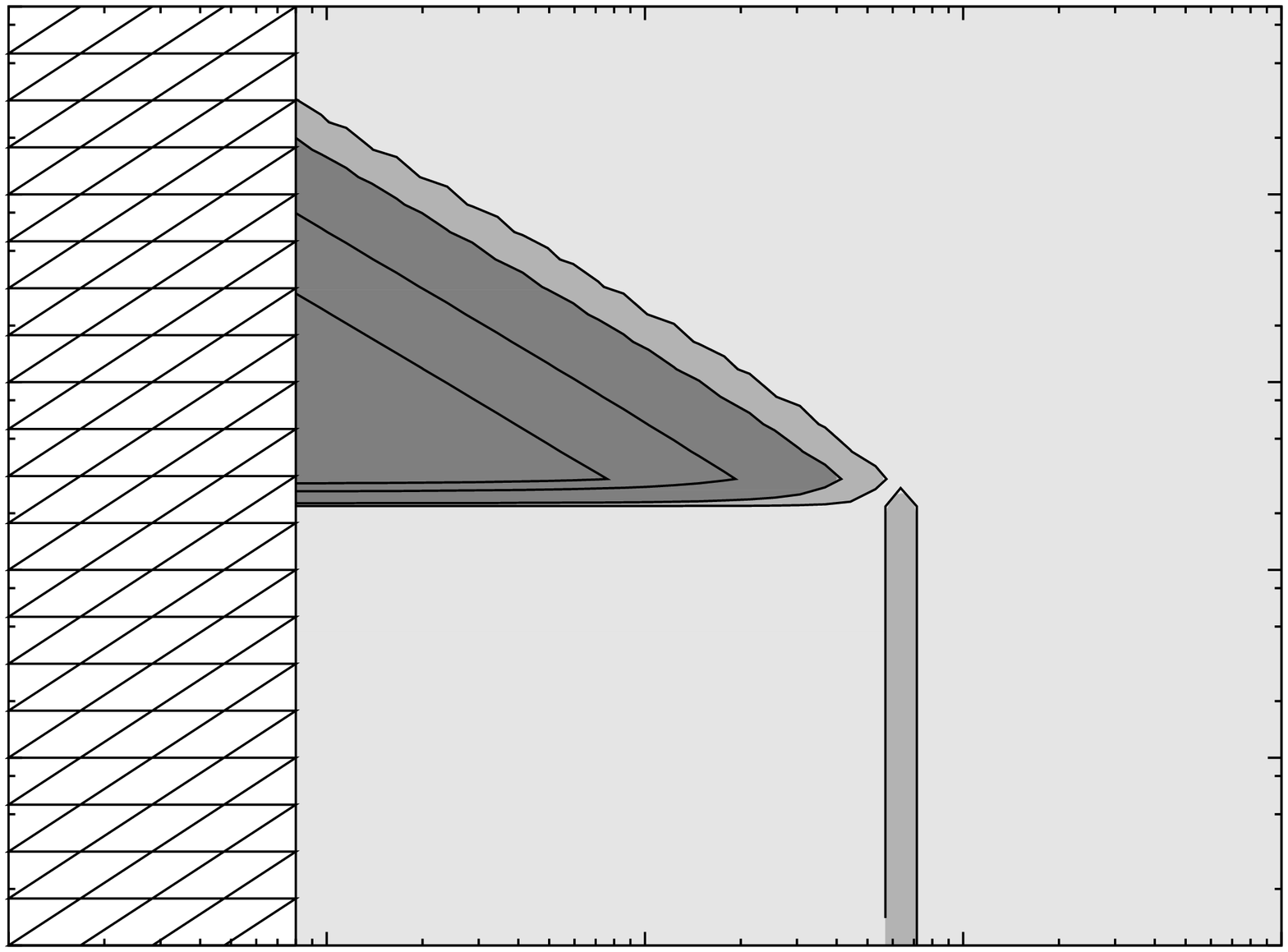} \\
\vspace{-0.3cm}
\includegraphics[ width=0.25\textwidth,angle=270]{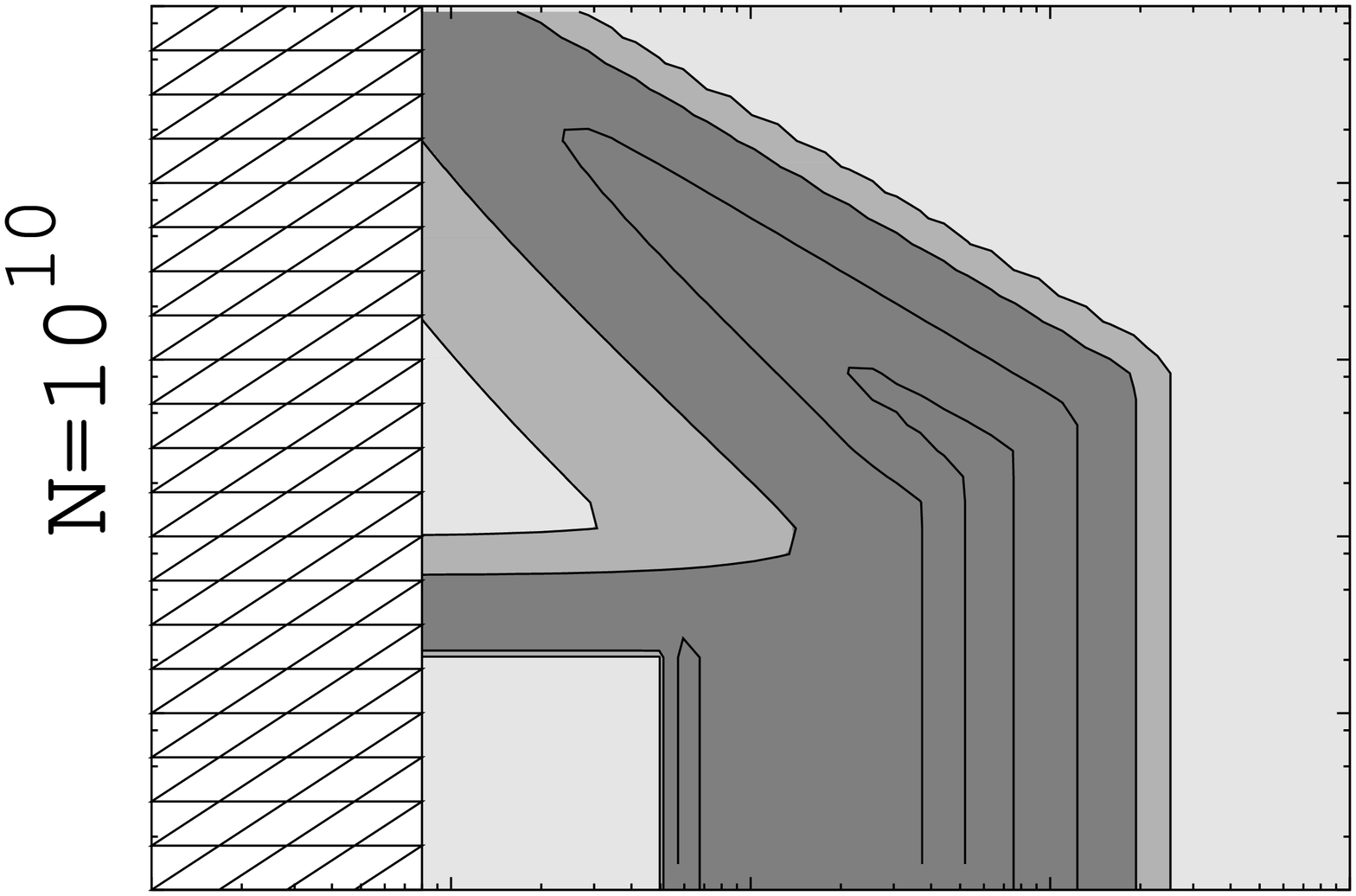} \hspace{-1cm}
\includegraphics[ width=0.25\textwidth,angle=270]{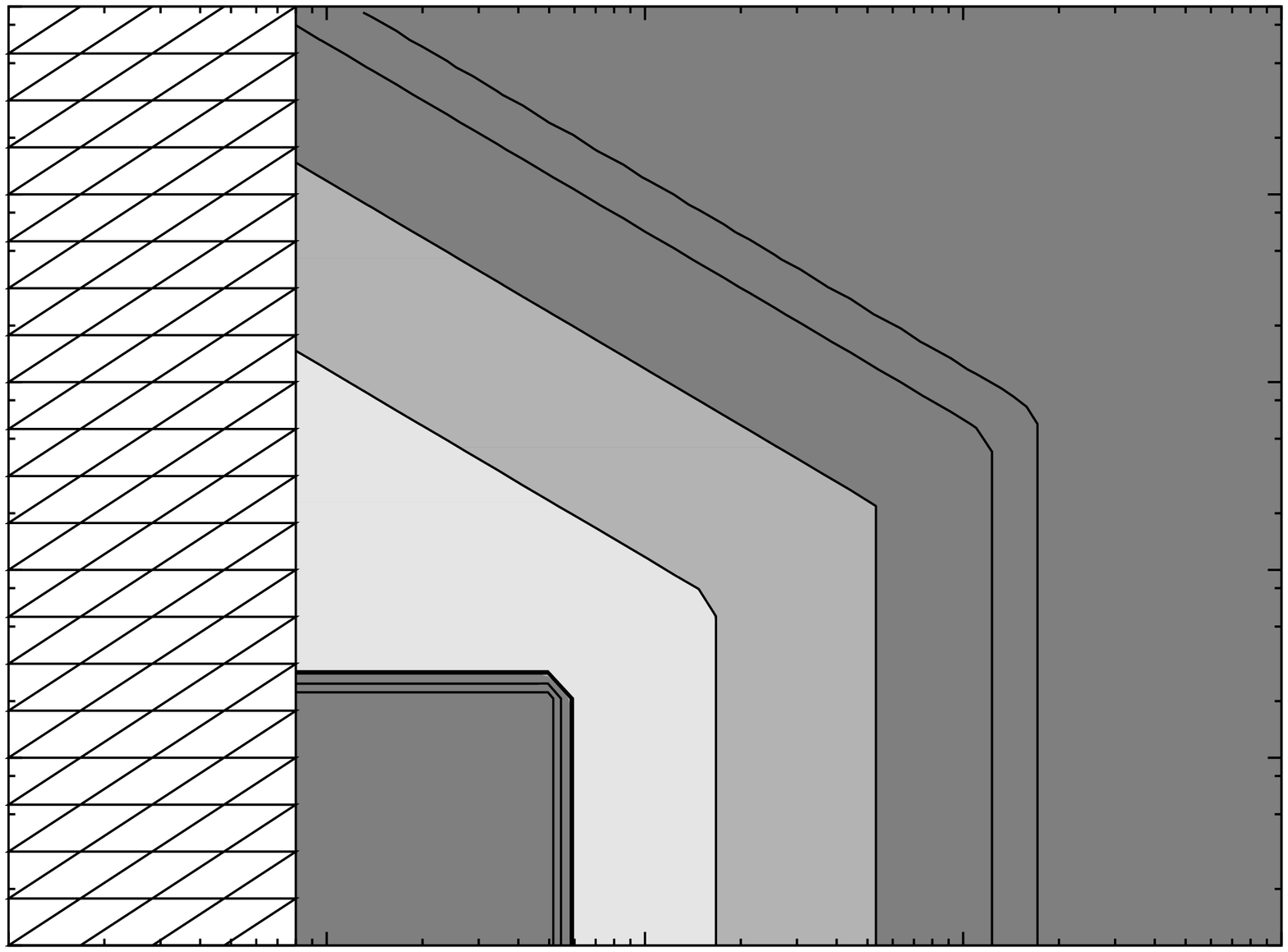}
\hspace{-0.95cm}
\includegraphics[ width=0.25\textwidth,angle=270]{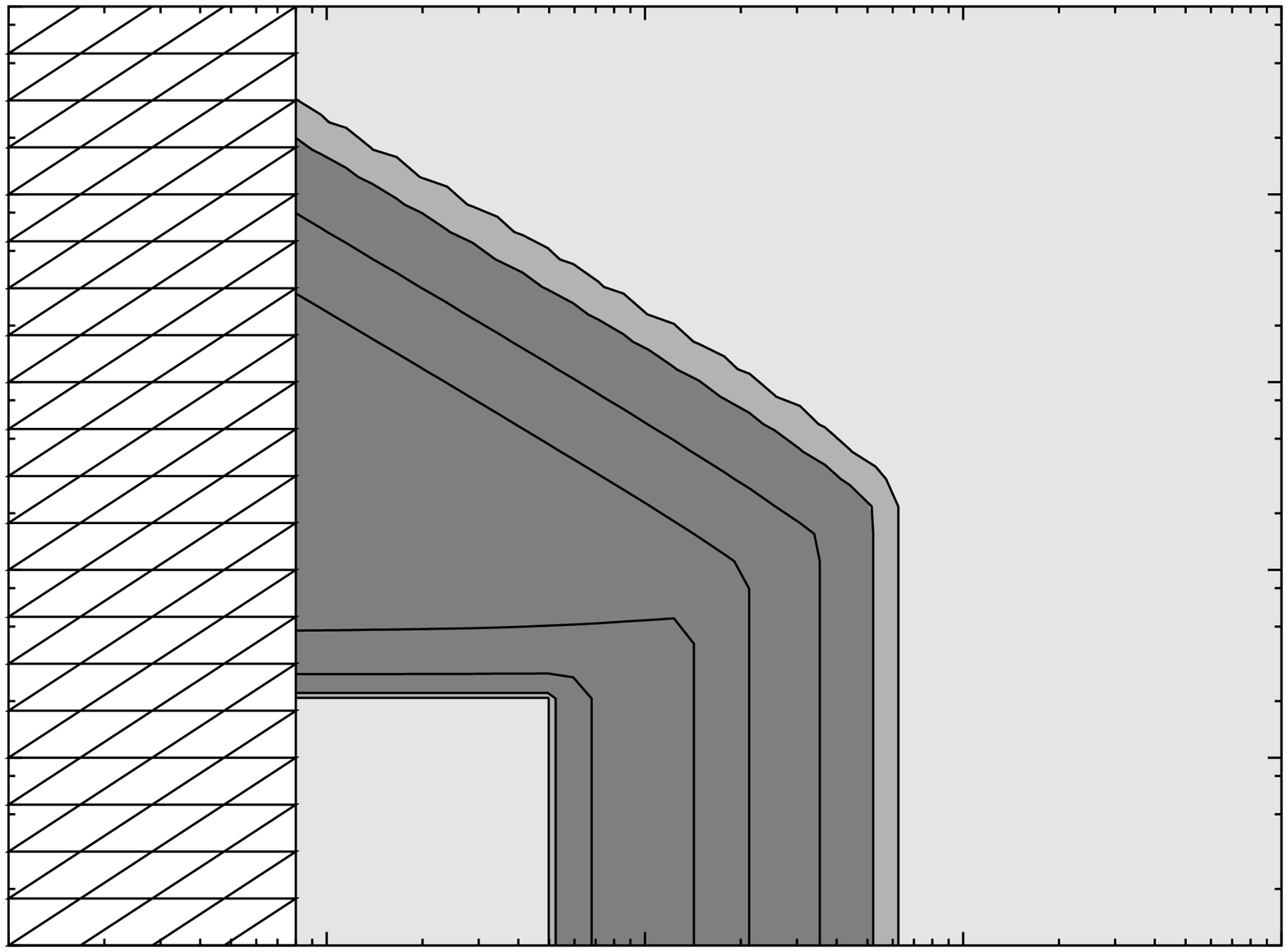} \\
\vspace{-0.3cm}
\includegraphics[ width=0.25\textwidth,angle=270]{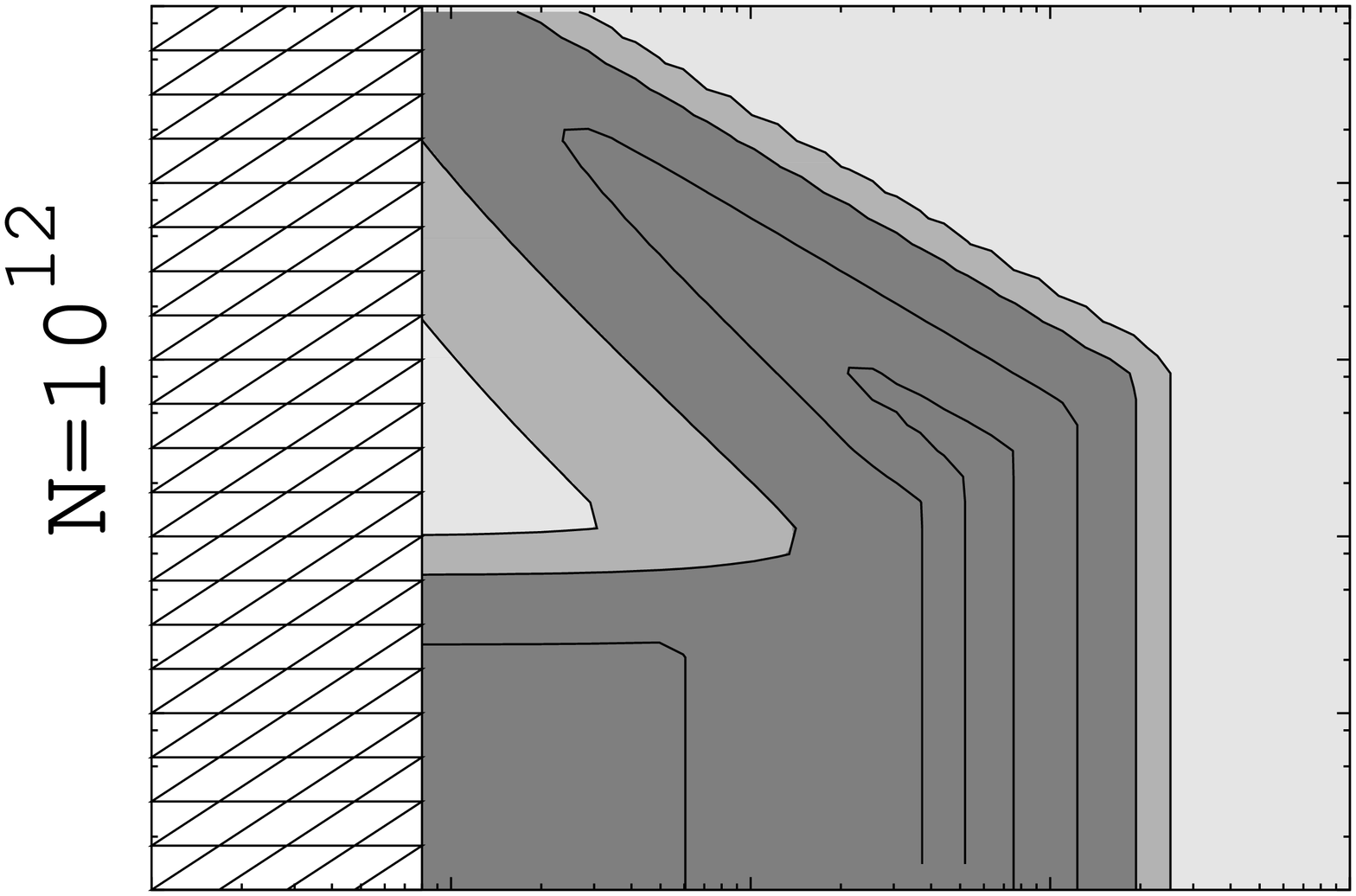} \hspace{-1cm}
\includegraphics[ width=0.25\textwidth,angle=270]{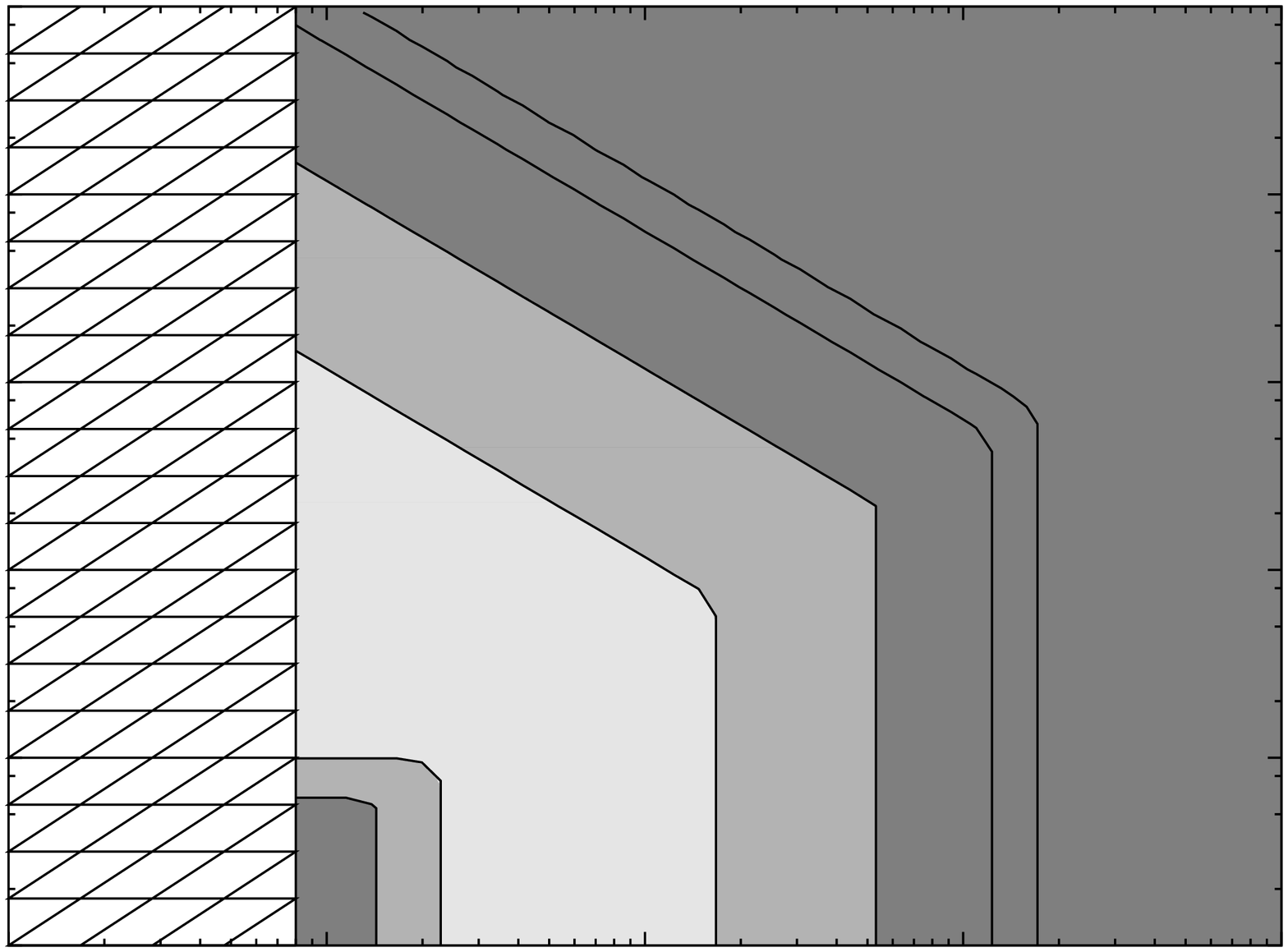}
\hspace{-0.95cm}
\includegraphics[ width=0.25\textwidth,angle=270]{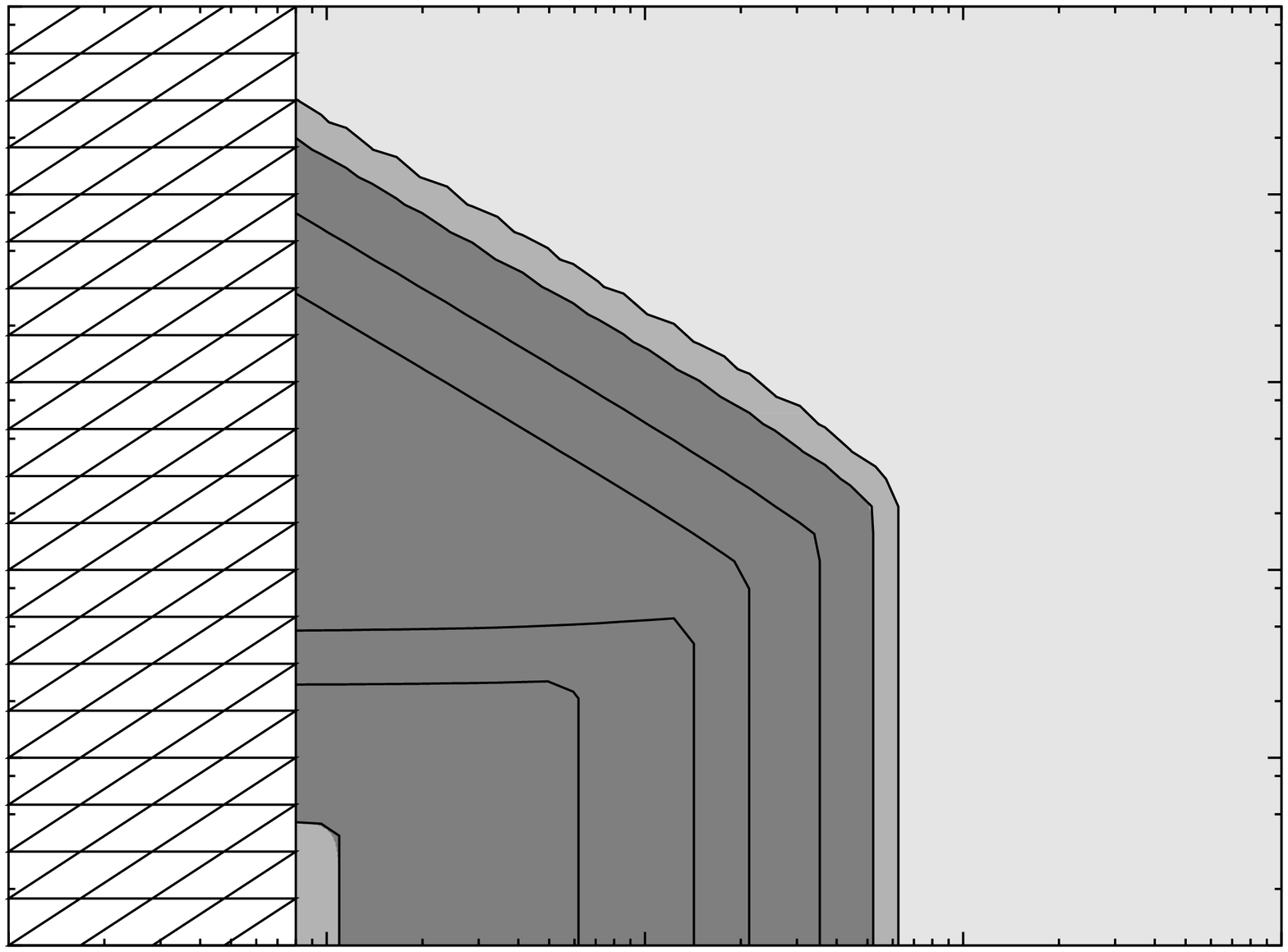} 
\caption{\label{fig:N_sigMpl}
$\sigma_0 = \mpl$: regions \obs,\ \negl\ and \excl\ as defined in the text, for increasing $N_{pre} = 10^4, 10^6, 10^8, 10^{10}, 10^{12}$. Axes, shading and parameters identical to \fig{fig:higgscont} and \fig{fig:Nsig_0}, but note different values of $\Npre$ compared to \fig{fig:Nsig_0}.}
\end{center}
\vspace{-0.3cm}
\end{figure}

\fig{fig:Nsig_0} shows the change in the three regions as $N_{pre}$ increases, for $\sigma_0 = 0$. Although \negl\ dominates at low $\Npre$, \obs\ is non-zero at small $m_\sigma$ even for $N_{pre}=10$. The distribution slowly reaches the equilibrium case, where small $m_{eff}$ (i.e.\ small $m_\sigma$ and small $g$) reach this last. Important results are that the large $m_\sigma$ region is {\em always} negligible for any $\Npre$, and that there is always some \obs\ region for small $m_\sigma$.

There is a non-trivial relationship between the curvaton parameters and the total duration of inflation. For example $m_\sigma = 10^{7}\GeV$ requires more than $10^4$ $e$-folds of inflation to become ``probable''. If the duration of inflation were to be constrained by some other method, then we gain information on the probable curvaton mass: small $N_{pre}$ corresponds to small $m_\sigma$; large $N_{pre}$ corresponds to a range of $m_\sigma$; very large $m_\sigma$ is never favoured. If instead we discover indications of the curvaton's mass, for example from CMB measurements of the primordial power spectrum\footnote{In general this requires information about the inflaton potential.}, then we can place constraints on the probable duration of inflation before the final $N_{obs}$ $e$-folds.

Although a symmetry argument might be invoked to motivate $\sigma_0 = 0$, it could also take another value. For $\sigma_0 = M_P$, the distribution again approaches the equilibrium beginning with large $m_{eff}$. This is shown in \fig{fig:N_sigMpl}. In contrast to the $\sigma_0 = 0$ case, \negl\ dominates entirely until $N_{pre}\simeq 10^4$. For larger $N_{pre}$, the distribution closely matches the equilibrium distribution for $N_{pre}>N_{rel}$, i.e.\ for large enough $m_{eff}$. Thus, the last region to reach the equilibrium distribution is a rectangle in the bottom left corner of the plot. The evolution in this case is somewhat simpler, because there is no overlap between the values of $\sigma_*$ needed to match observations and the distribution of $\sigma_*$, until the central value has moved sufficiently away from $\mpl$. In the $\sigma_0 = 0$ case, the central value was initially at the equilibrium value.

If a fundamental theory were to have set $\sigma_0 = \mpl$ before inflation, then knowledge of $g$ and/or $m_\sigma$ could give information about the duration of inflation. For example, if $N_{pre}$ was known to be small, the curvaton model would be disfavoured.

Although we have carried out this analysis for a particular realisation of a curvaton model (the MCH model) with particular choices of $\sigma_0$, the analysis could easily be repeated for other scenarios. Thus, our method would enable the likely effects of an additional scalar e.g.\ in a supersymmetric inflation model to be determined. Under some general assumptions it would be possible to determine which masses and couplings would lead to an observable effect on the scalar perturbations. Thus, this methodology may have  interesting implications for many different models.

\section{Conclusions}
\label{sec:conc}

We can think about spectator fields during inflation in two ways. The first is that if (non-inflaton) scalar fields exist in a theory, do they either rule out the theory or otherwise affect observational predictions? The second is that if we design a model such that a spectator field is responsible for the observed density perturbation, does this occur with natural or with fine-tuned initial conditions? This paper provides a framework to answer both of these questions. In the absence of fine-tuned initial conditions, we find in almost all cases that probable regions of parameter space exist for some range of model parameters (considering the minimal curvaton-higgs (MCH) model \cite{Enqvist:2013gwf} with $H_* = 10^{10}\GeV$). In addition, large masses $m_\sigma\gtrsim 2\times 10^7\GeV$ are disfavoured due to negligible curvature perturbations.

We obtained our results by assuming the curvaton had the initial condition $\sigma_0 = 0$ (\fig{fig:Nsig_0}) or $\sigma_0 = \mpl$ (\fig{fig:N_sigMpl}) before inflation. We assumed that inflation lasted $N = N_{pre} + N_{obs}$ $e$-folds and considered various values of $N_{pre}$ including the $N_{pre}\to \infty$ limit (\fig{fig:higgscont}). From the initial strongly peaked $\sigma_0$, the long wavelength modes of the curvaton field behave stochastically. We calculated the probability distribution for the mean field value $\sigma$ at a given time; the observable Universe is given by one particular realisation of this distribution. We then translated this probability distribution to a probability of obtaining particular ranges for the curvature perturbation $\zeta$ and the non-Gaussianity $\fnl$. In this way we attempted to quantify how likely a set of parameters are to give an \obs,\ \negl\ or \excl\ Universe.

We used the MCH model \cite{Enqvist:2013gwf} throughout the paper to illustrate our method. In this model the curvaton has a quadratic potential and is coupled only to the standard model Higgs Boson with coupling $g$. For simplicity, we ignored a number of effects, including the effective quartic coupling of the curvaton and the possibility of an inflaton-curvaton coupling. We also assumed that the dominant mechanism for curvaton decay was through dimension-5 non-renormalisable couplings. At $g\lesssim (m_\sigma/\mpl)^{1/4}$ this assumption is likely to be reasonable, but the behaviour of such an oscillating field in a thermal background after inflation is complicated and is the subject of ongoing study  (see \cite{D'Onofrio:2012qy,Enqvist:2013qba,BasteroGil,Drewes:2013iaa,Mukaida:2014yia,Mukaida,Enqvist:2012tc}). For example, the generation of perturbations in a case similar to the MCH model could also involve ``self-modulated preheating'' \cite{Mukaida:2014yia}. Including all of these effects in our calculations would certainly be interesting, although it would not be possible within our semi-analytical framework.

We found that $m_\sigma > 2\times 10^{7} \GeV$ always gives a Universe with negligible $\zeta$ with $>99\%$ probability. Masses below $m_\sigma = 8\times 10^4\GeV$ are excluded due to a late curvaton decay spoiling the predictions of BBN. Considering $8\times 10^{4}\GeV \leq m_\sigma \leq 2\times 10^7\GeV$, observable $\zeta$ has high probability ($> 10\%$ and even $> 50\%$) for broad ranges of $m_\sigma$ and $g$. A notable region with small $m_\sigma$ and large $g$  is likely to give a too large curvature perturbation. There is very little dependence on the initial conditions for large effective mass $m_{eff}^2 = m_\sigma^2 + g^2 h_*^2$ because the equilibrium distribution is already achieved for small $\Npre$. For smaller $m_{eff}$ (small $m_\sigma$ and $g$), there is sensitivity to the initial field value $\sigma_0$ and to the duration of inflation. In general, larger values of $m_\sigma$ become probable only after a longer duration of inflation. All of these conclusions are for $H_* = 10^{10}\GeV$.

Future experiments, observations and theoretical progress are expected to further constrain the MCH model. For example, the Planck satellite will be sensitive to non-Gaussianity down to $\fnl \simeq 5$ and can also constrain $H_*$. Further theoretical calculations of curvaton decay mechanisms are likely to constrain $m_\sigma$ and $g$. After an inflation model is assumed, $m_\sigma$ and $g$ may also be constrained by measurements of the CMB spectral index.  Additionally, if WIMP dark matter were to be favoured, the curvaton would be required to decay before dark matter freeze-out to avoid large isocurvature --- and because the dimension-5 decay width depends on $m_\sigma$, a large part of the low $m_\sigma$ parameter space would then be ruled out. Together, these advances may make it possible to link the MCH model with the duration of inflation. For example if the above considerations favoured large $m_\sigma$, a curvaton model without fine-tuning would require a long duration of inflation. If small $m_\sigma$ and large $g$ was instead favoured, $N_{pre} = {\cal O}(10-100)$ would be sufficient for a ``probable'' Universe, provided that $\sigma_0 = 0$ could be justified. Conversely, knowledge from fundamental theories about the total duration of inflation could also give information about the parameters in the curvaton model.

In summary, we have studied the ``naturalness'' of the MCH scenario under generic assumptions about the curvaton initial conditions and various other simplifications. We find that for $8\times 10^4\GeV \leq m_\sigma \leq 2\times 10^7\GeV$, the curvaton mechanism can ``naturally'' occur, provided that inflation has lasted long enough. Our methodology could be applied to more general curvaton models, as well as to models with additional spectator fields. It could then be determined whether these additional fields must have fine-tuned initial conditions to avoid spoiling the predictions of the model.

\section*{Acknowledgements}
RL and SM were supported by the Academy of Finland grant 1263714. We thank Kari Enqvist for enlightening discussions, and Wilfried {Buchm\"{u}ller} for insightful comments on the final manuscript.

\end{document}